\documentclass[fleqn,usenatbib]{mnras}

\usepackage{newtxtext,newtxmath}
\usepackage{subcaption}
\usepackage{multirow}
\usepackage{wasysym}

\usepackage[T1]{fontenc}

\DeclareRobustCommand{\VAN}[3]{#2}
\let\VANthebibliography\thebibliography
\def\thebibliography{\DeclareRobustCommand{\VAN}[3]{##3}\VANthebibliography}

\usepackage{graphicx}	% Including figure files
\usepackage{amsmath}	% Advanced maths commands

\usepackage{booktabs}

\title[Evidence for non-circular motions and AGN feedback in GN20]{GA-NIFS: NIRSpec reveals evidence for non-circular motions and AGN feedback in GN20}

\author[H. {\"U}bler et al.]{\parbox{\textwidth}{
Hannah {\"U}bler,$^{1,2}$\thanks{hu215@cam.ac.uk}
Francesco D'Eugenio,$^{1,2}$
Michele Perna,$^{3}$
Santiago Arribas,$^{3}$ 
Gareth C.~Jones,$^{4}$
Andrew J.~Bunker,$^{4}$  
Stefano Carniani,$^{5}$
St{\'e}phane Charlot,$^{6}$ 
Roberto Maiolino,$^{1,2,7}$ 
Bruno Rodr{\'i}guez del Pino,$^{3}$ 
Chris J.~Willott,$^{8}$  
Torsten B{\"o}ker,$^{9}$ 
Giovanni Cresci,$^{10}$
Nimisha Kumari,$^{11}$
Isabella Lamperti,$^{3}$
Eleonora Parlanti,$^{5}$ 
Jan Scholtz,$^{1,2}$
Giacomo Venturi$^{5}$
\\
}
\\
\parbox{\textwidth}{
$^{1}$Kavli Institute for Cosmology, University of Cambridge, Madingley Road, Cambridge CB3 0HA, UK\\
$^{2}$Cavendish Laboratory, University of Cambridge, 19 JJ Thomson Avenue, Cambridge, CB3 0HA, UK\\
$^{3}$Centro de Astrobiolog\'{\i}a (CAB), CSIC-INTA, Ctra. de Ajalvir km 4, Torrej\'on de Ardoz, E-28850, Madrid, Spain\\
$^{4}$University of Oxford, Department of Physics, Denys Wilkinson Building, Keble Road, Oxford OX13RH, UK\\
$^{5}$Scuola Normale Superiore, Piazza dei Cavalieri 7, I-56126 Pisa, Italy\\
$^{6}$Sorbonne Universit\'e, CNRS, UMR 7095, Institut d' Astrophysique de Paris, 98 bis bd Arago, 75014 Paris, France\\
$^{7}$Department of Physics and Astronomy, University College London, Gower Street, London WC1E 6BT, UK\\
$^{8}$National Research Council of Canada, Herzberg Astronomy \& Astrophysics Research Centre, 5071 West Saanich Road, Victoria, BC V9E 2E7, Canada\\
$^{9}$European Space Agency, c/o STScI, 3700 San Martin Drive, Baltimore, MD 21218, USA\\
$^{10}$INAF - Osservatorio Astrofisco di Arcetri, largo E. Fermi 5, 50127 Firenze, Italy\\
$^{11}$AURA for the European Space Agency, Space Telescope Science Institute, Baltimore, MD, USA
 }
}

\date{Accepted XXX. Received YYY; in original form ZZZ}

\pubyear{2024}

\begin{document}
\label{firstpage}
\pagerange{\pageref{firstpage}--\pageref{lastpage}}
\maketitle

\begin{abstract}
We present rest-frame optical data of the $z\sim4$ sub-millimeter galaxy GN20 obtained with {\it JWST}/NIRSpec in integral field spectroscopy (IFS) mode. The H$\alpha$ emission is asymmetric and clumpy and extends over a projected distance of more than 15~kpc. To first order, the large-scale ionised gas kinematics are consistent with a turbulent ($\sigma\sim90$~km/s), rotating disc ($v_{\rm rot}\sim500$~km/s), congruent with previous studies of its molecular and ionised gas kinematics. However, we also find clear evidence for non-circular motions in the H$\alpha$ kinematics. We discuss their possible connection with various scenarios, such as external perturbations, accretion or radial flows. 
In the centre of GN20, we find broad line emission (FWHM~$\sim1000-2000$~km/s) in the H$\alpha$+[N~II] complex, suggestive of fast, AGN-driven winds or, alternatively, of the broad-line region of an active black hole.
Elevated values of [N{\sc ii}]$\lambda6583$/H$\alpha>0.4$ and EW(H$\alpha)>6$~\AA\, throughout large parts of GN20 suggest that feedback from the active black hole is able to photo-ionise the interstellar medium. Our data corroborates that GN20 offers a unique opportunity to observe key processes in the evolution of the most massive present-day galaxies acting in concert, over 12 billion years ago. 
\end{abstract}

\begin{keywords}
galaxies: kinematics and dynamics -- galaxies: active -- galaxies: high-redshift
\end{keywords}

%%%%%%%%%%%%%%%%%%%%%%%%%%%%%%%%%%%%%%%%%%%%%%%%%%

%%%%%%%%%%%%%%%%% BODY OF PAPER %%%%%%%%%%%%%%%%%%

\section{Introduction}

More than half of the total stellar mass in the present-day Universe is situated in early-type galaxies \citep[ETGs; ][]{Fukugita98, Hogg02, Bell03, Renzini06}. Understanding the formation of these giant systems is one of the key goals of galaxy evolution studies. With estimated ETG formation redshifts of $z>3$, the discovery of bright sub-millimeter galaxies \citep[SMGs, see review by][]{Blain02} at $z\gtrsim4$ quickly promoted them to candidate progenitors \citep[e.g.][]{Simpson14}. These galaxies rapidly assemble high stellar masses through intense star formation \citep[e.g.][]{Hainline09, Magnelli13, Casey14, Swinbank14}. Besides, some SMGs, including the galaxy studied in this work, are located within cosmic over-densities, hinting at future merger events which would help them to further increase their mass by $z=0$ \citep[e.g.][]{Daddi09, Walter12, Casey14, Oteo18, Pavesi18, AlvarezMarquez23, Arribas23, Jones23}. 
Furthermore, at redshifts $z>3$ the cores of proto-clusters, which are believed to reside at the nodes of the cosmic web, are expected to be subject to cold gas inflows \citep[e.g.][]{Overzier16}, a process scarcely studied observationally.

After the identification of many $z>4$ SMGs in the past decades \citep[e.g.][]{Smail97, Barger98, Hughes98, Blain02, Chapman03, Chapman05, Pope05, Simpson14, Hodge20}, the advent of the {\it James Webb Space Telescope} ({\it JWST}) has enabled, for the first time, studies of their rest-frame optical emission-line properties. 
Within the GA-NIFS survey\footnote{\url{https://ga-nifs.github.io}} (`Galaxy Assembly with NIRSpec IFS'; PIs: Santiago Arribas, Roberto Maiolino), as part of the NIRSpec Instrument Science Team Guaranteed Time Observations, we have targeted several bright SMGs at $z\geq4$ with the integral-field spectroscopic (IFS) mode \citep{Jakobsen22, Boeker22}, three of which have already been discussed: ALESS073.1 at $z=4.76$, for which the NIRSpec-IFS data reveal a heavily dust-obscured active galactic nucleus (AGN) \citep{Parlanti23}; HLFS3 at $z=6.3$, which \cite{Jones23} identify as a dense galaxy group in the process of merging; and the massive proto-cluster core SPT0311-58 at $z=6.9$, in which \cite{Arribas23} find evidence for accretion from the cosmic web, inflows and mergers.
In this paper, we present high-resolution ($R\sim2700$) NIRSpec-IFS data of the rest-frame optical line-emission in GN20, a large SMG at $z\sim4$. 

GN20 was detected by \cite{Pope05} as a bright $850~\mu$m source in the Great Observatories Origins Deep Survey Northern (GOODS-N) field \citep[see also][]{Iono06, Pope06}.
This dusty star-forming galaxy is located within a proto-cluster environment, and is detected in 1.4~GHz continuum, [C~I] and in several CO transitions (1-0, 2-1, 4-3, 5-4, 6-5, 7-6) (\citealp{Daddi09, Carilli10, Morrison10, Hodge13, Cortzen20}; see also \citealp{Casey09}). 
It has a total infrared luminosity of $\sim2-3\times10^{13}~L_\odot$, a star-formation rate of SFR$_{\rm IR}\sim1800-3000~M_\odot$/yr, a stellar mass of $M_\star\sim1.1-2.3\times10^{11}~M_\odot$, and a molecular gas mass of $M_{\rm mol}\sim5-13\times10^{10}~M_\odot$ \citep{Daddi09, Carilli10, Hodge12, Tan14}. 
A regular velocity gradient spanning about 570~km/s is observed in CO(4-3), consistent with an exceptionally large $r\sim4$~kpc and massive rotating disc \citep{Carilli10}. \cite{Hodge12} find a clumpy, rotating disc in CO(2-1) with a maximum rotation velocity of $v_{\rm rot,max}=575\pm100$~km/s, with a velocity dispersion of $\sigma=100\pm30$~km/s.
Dynamical mass estimates for GN20 are in the range $2-6\times10^{11}~M_\odot$ \citep{Daddi09, Carilli10, Carilli11, Hodge12}.
Intriguingly, rest-frame ultra-violet (UV) emission tracing young stars as observed with HST/WFC3 F105W imaging is visible only in an extended ($\sim9$~kpc) stripe to the North-West of the centroids of the CO and 880~$\mu$m emission, beyond the extent of the dust emission \citep{Hodge15}.

Recently, \cite{Colina23} studied the rest-frame 1.1~$\mu$m imaging of GN20 obtained with the MIRI instrument onboard {\it JWST} \citep[see also][]{CrespoGomez24}. Their analysis reveals a two-component stellar structure composed of an unresolved nucleus offset by 1~kpc from the centre of an extended disc with $R_e=3.60$~kpc. 
They argue that the offset nucleus may be a result of tidal interactions with other proto-cluster members, or indicate a late-stage merger. \cite{Colina23} find that the stellar nucleus coincides with the centre of far-infrared continuum emission tracing dust-obscured star formation, while the extended stellar envelope overlaps with the cold molecular gas distribution. 
Analysing MIRI/MRS-IFU observations, \cite{Bik23} find clumpy Pa$\alpha$ emission out to a radius of 6~kpc. The kinematics of the Pa$\alpha$ emission are consistent with a rotating disc, with a maximum rotation velocity of $v_{\rm rot}=550\pm40$~km/s and an upper limit on the flux-weighted velocity dispersion ($\sigma_m$) of $\sigma_m=145\pm53$~km/s. 
Comparing the unobscured star-formation rate derived from the integrated Pa$\alpha$ flux, SFR$_{\rm Pa\alpha}=144\pm9~M_\odot$/yr, to the infrared star-formation rate SFR$_{\rm IR}$ \citep{Tan14}, \cite{Bik23} infer a high average extinction of A$_V=17.2\pm0.4$~mag \citep[see also][]{Maseda24}. 
\cite{CrespoGomez24} infer a lower A$_V\sim1.5$~mag from spectral energy distribution (SED) fitting. They attribute this difference to the presence of either stellar populations older than 10~Myr, or of a buried AGN.

We describe the {\it JWST}/NIRSpec-IFS high-resolution observations of GN20 and our analysis methods in Sections~\ref{s:obs} and \ref{s:fitting}. We present the complex H$\alpha$ emission revealed by NIRSpec in Section~\ref{s:hamorph}. In Section~\ref{s:kinematics} we discuss the H$\alpha$ kinematics and evidence for non-circular motions based on both our data and dynamical modelling. In Section~\ref{s:fuel} we show evidence for the presence of an AGN in the centre of GN20. We summarise our findings in Section~\ref{s:conclusions}.

Throughout this work, we assume a flat $\Lambda$CDM cosmology with $\Omega_m=0.315$ and $H_0=67.4$ km/s/Mpc \citep{Planck20}.  With this cosmology, $1''$ corresponds to a transverse distance of 7.07 proper kpc at $z=4.05$.

\section{NIRSpec-IFU observations and Data Processing}\label{s:obs}

GN20 was observed in NIRSpec-IFS mode as part of the GA-NIFS survey under programme 1264. The NIRSpec data were taken on February 10, 2023, with a medium cycling pattern of four dither positions and a total integration time of about 2~h with the high-resolution grating/filter pair G395H/F290LP, covering the wavelength range $2.87-5.14\mu$m (spectral resolution $R\sim2000-3500$; \citealp{Jakobsen22}), and about 1~h with PRISM/CLEAR ($\lambda=0.6-5.3\mu$m, spectral resolution $R\sim30-300$).
Within the same programme, MIRI \citep{Rieke15, Wright15} imaging and spectroscopy of GN20 was obtained. The MIRI imaging data was presented by \cite{Colina23}, \cite{CrespoGomez24}, and the MRS data by \cite{Bik23}.

Raw data files were downloaded from the Barbara A.~Mikulski Archive for Space Telescopes (MAST) and subsequently processed with the {\it JWST} Science Calibration pipeline\footnote{\url{https://jwst-pipeline.readthedocs.io/en/stable/jwst/introduction.html}} version 1.11.1 under the Calibration Reference Data System (CRDS) context jwst\_1149.pmap. We made several modifications to the default reduction steps to increase data quality, which are described in detail by \cite{Perna23} and which we briefly summarise here. 
Count-rate frames were corrected for $1/f$ noise through a polynomial fit. 
During calibration in Stage 2, we removed regions affected by failed open MSA shutters. We also removed regions with strong cosmic ray residuals in several exposures.
Remaining outliers were flagged in individual exposures using an algorithm similar to {\sc lacosmic} \citep{vDokkum01}: we calculated the derivative of the count-rate maps along the dispersion direction, normalised it by the local flux (or by three times the rms noise, whichever was highest), and rejected the 95\textsuperscript{th} percentile of the resulting distribution \citep[see][for details]{DEugenio23}. 
The final cube was combined using the `drizzle' method. The main analysis in this paper is based on the combined cube with a pixel scale of $0.05''$.
We used spaxels away from the central source and free of emission features to perform a background subtraction.

\begin{figure}
    \centering
   \includegraphics[width=\columnwidth]{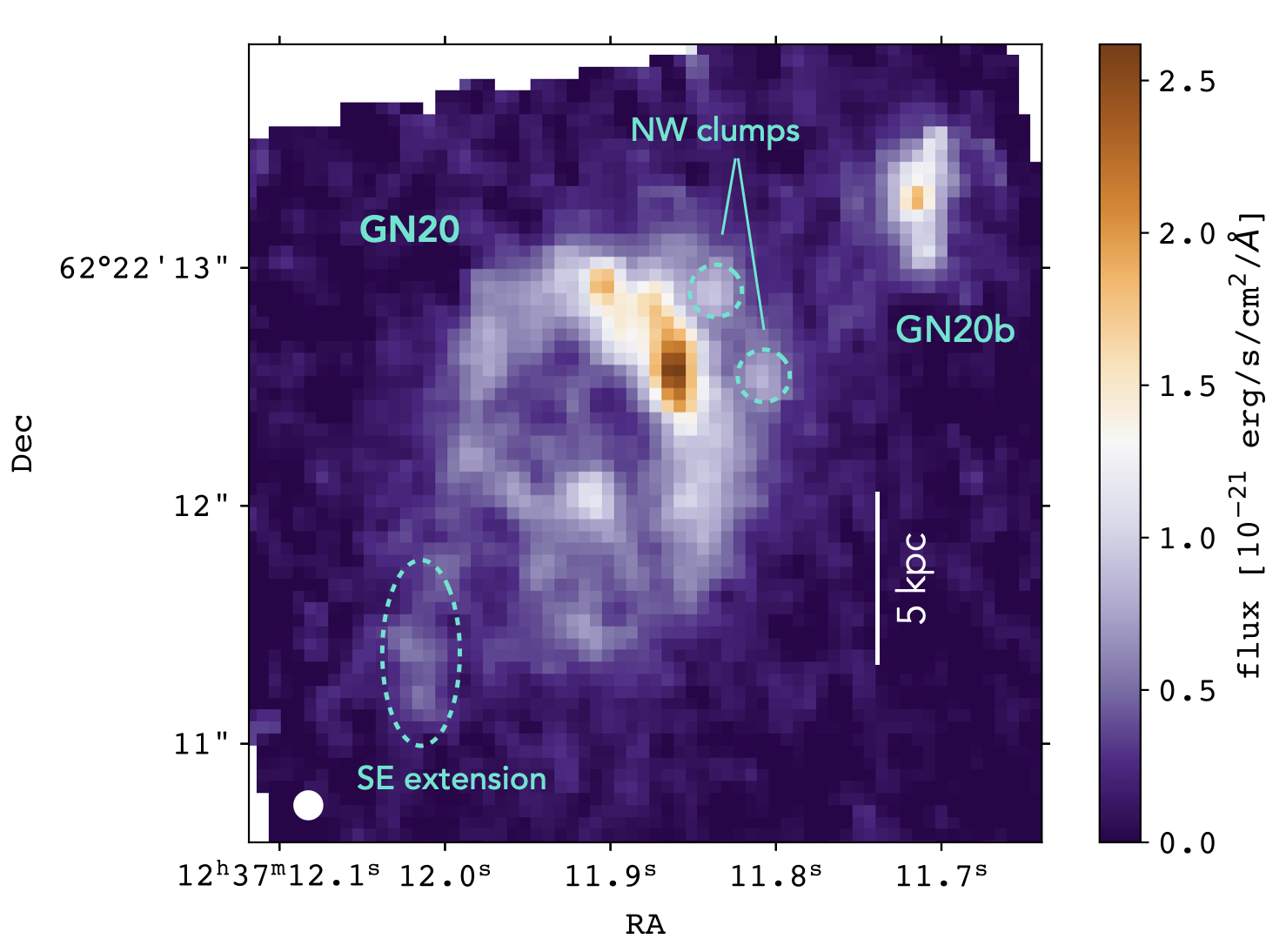}
    \caption{Line map of the emission integrated in the range $3.313-3.327\mu$m, roughly encompassing the H$\alpha$+[N~II]$\lambda$6583 lines at $z\sim4.055$. GN20 is visible as the extended central source with a complex and clumpy morphology. We indicate two clumps to the North-West and extended emission in the South-East with dashed outlines (see main text for details). We detect a second galaxy to the North-West of GN20, which we call GN20b, at a projected separation of about 12~kpc and a velocity difference of about +750~km/s, respectively a redshift of $z=4.06532$. The white circle indicates the approximate PSF FWHM ($\diameter=0.12''$).}
    \label{f:halinemap}
\end{figure}

In Figure~\ref{f:halinemap}, we show a line map of the emission integrated in the range $3.313-3.327\mu$m, roughly encompassing the H$\alpha$ and [N~II]$\lambda6583$ lines at $z\sim4.055$. This will be further discussed in Section~\ref{s:hamorph}.
In this paper, we focus on the analysis of the high-resolution data. However, to provide some context for our analysis, in Figure~\ref{f:rgb} we use the prism observations to create a false-colour multi-wavelength image of GN20, combining the H$\alpha$+[N~II]$\lambda\lambda6548,6583$ emission (green) with the emission at $0.9-1.2\mu$m (blue), roughly corresponding to the {\it HST}/WFC3 F105W filter tracing the emission of unobscured young stars, and the emission at $5.0-5.3\mu$m (red), covering the blue part of the MIRI/F560W filter and tracing stellar mass.
The astrometry in Figure~\ref{f:halinemap} is derived by registering our observations to the F560W and F105W images, which are in turn registered to Gaia DR3 \citep{GaiaDR3}. Based on the positional shifts derived from the MIRI and HST data, and the varying pixel sizes of the observations ranging from 0.05” to 0.1”, we estimate an uncertainty of about 0.1” on the astrometry.

\section{Fitting and Dynamical Modelling}\label{s:fitting}

\subsection{Emission line fitting and H$\alpha$ redshift}

To analyse the emission-line properties and kinematics of GN20, we first fit a one-component Gaussian model to the H$\alpha$ and [N{\sc ii}]$\lambda\lambda6548,6583$ emission lines in the grating data, including a constant continuum. 
Because the signal-to-noise ratio ($S/N$) varies substantially throughout GN20 due to non-uniform dust obscuration that is particularly high in the central region \citep{Hodge15, Colina23, Bik23}, we derive a Voronoi-binned map, using the algorithm \textsc{vorbin} \citep{Cappellari03}, as implemented in \textsc{QFitsView}. We first define a mask based on visual inspection of the line emission in the cube, and then require $S/N\geq15$ in the previously fitted H$\alpha$ flux map to derive the Voronoi map.
We derive H$\alpha$ velocity and velocity dispersion maps by repeating our one-component Gaussian model fitting on the binned data. The resulting maps are shown in the top row of Figure~\ref{f:kinmaps}.

While the line emission is relatively narrow in the outer regions of GN20 ($\sigma_{\rm obs}\sim50-100$ km/s), the H$\alpha$+[N{\sc ii}] emission in the centre is broad. We identify the central regions with broad emission through visual inspection (see aperture indicated in the bottom panels of Figure~\ref{f:centralspec}, diameter of $\sim4$~kpc).
We fit the spectra extracted from the central regions based on the Voronoi tessellation with a set of narrow components plus a set of broad components for H$\alpha$ and [N{\sc ii}]$\lambda\lambda6548,6583$. Here, we require FWHM$_{\rm broad}>400~{\rm km/s}>$~FWHM$_{\rm narrow}$.
We stitch the narrow component results to our initial one-component maps. 
As can be seen from the example spectra in Figure~\ref{f:centralspec}, the narrow components in the central region are distinct from the underlying broad emission. It is therefore unlikely that we over-subtract emission associated with the narrow component through this procedure.
The resulting maps are shown in the bottom row of Figure~\ref{f:kinmaps}. In Section~\ref{s:fuel} we discuss alternative approaches to model the broad nuclear emission.

To measure a redshift for GN20 from the H$\alpha$ data, we rely on the kinematics, since the one-dimensional, integrated emission line profile is skewed by the bright, off-centre region in the North-West. After initially deriving the H$\alpha$ kinematics as described above assuming $z=4.055$ \citep{Daddi09}, we shift the systemic velocity such that we find comparable absolute maximum and minimum velocities, and the (one-component) dispersion peak coincides with $v_{\rm obs}\sim0$~km/s along the kinematic major axis. We adopt a velocity shift of +75~km/s. Given the median uncertainties on the fitted velocities of 11~km/s (including in the regions of minimum and maximum velocities), this corresponds to a redshift of $z_{\rm H\alpha}=4.05374\pm0.00075$.\footnote{
As described in the Section~\ref{s:dyn}, in our dynamical modelling we allow for an additional velocity shift of $\pm100$~km/s. Our fiducial model has a velocity shift of $+28\pm2$~km/s. This would correspond to a redshift of $z_{\rm H\alpha,model}=4.05325\pm0.00075$. However, we caution that the velocity shift can be degenerate with the adopted centre (fixed in our model). In addition, as discussed in Section~\ref{s:kinematics}, the model does not capture all kinematic features of the data.}

\begin{figure}
    \centering
    \includegraphics[width=0.75\columnwidth]{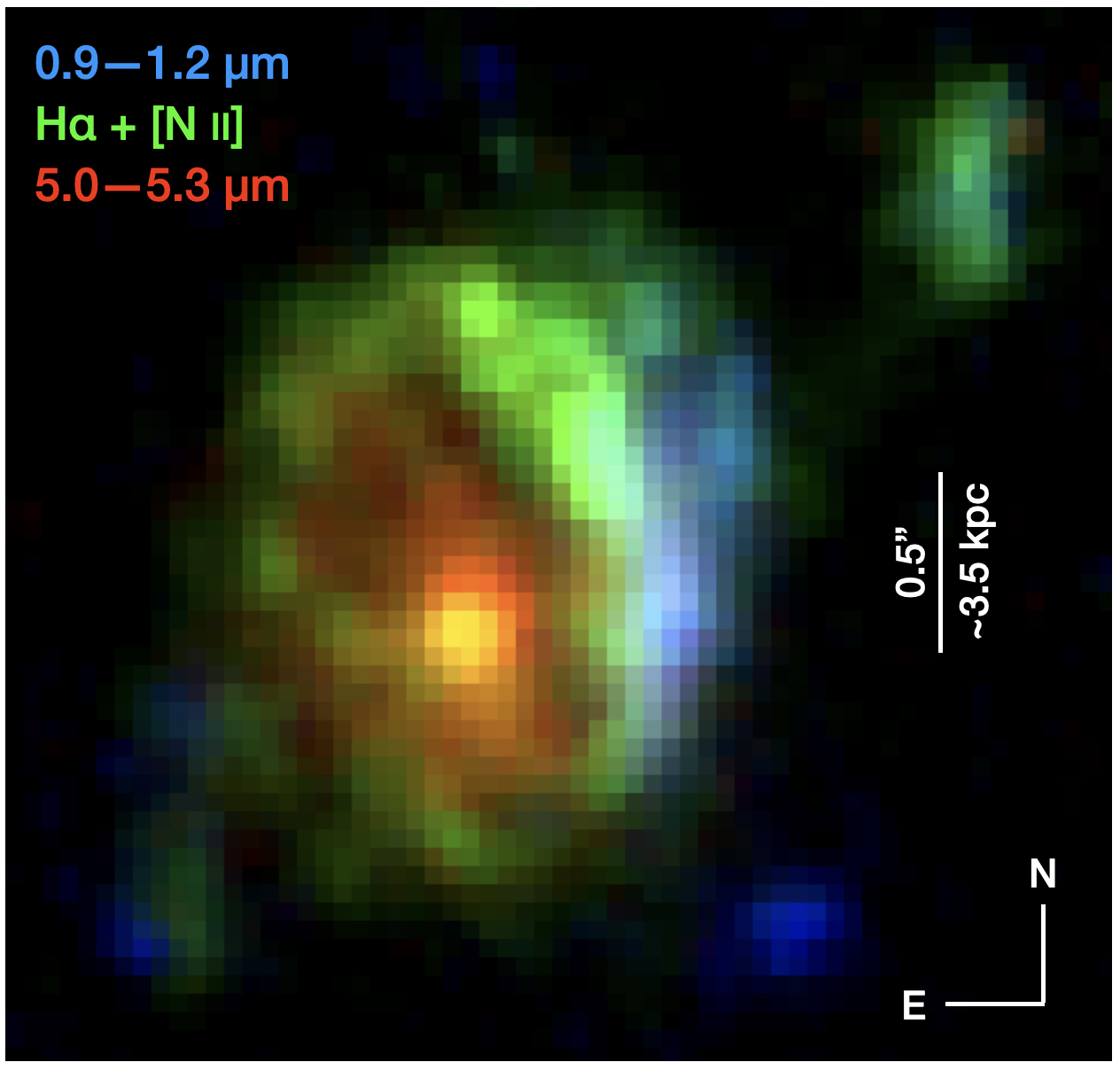}
    \caption{Comparison of rest-frame optical, UV, and near-infrared emission in GN20, obtained through collapsing the NIRSpec prism cube in three different wavelength regions (not PSF-matched). The H$\alpha$+[N~II] emission is shown in green, blue colors trace emission at $0.9-1.2\mu$m (rest-frame $0.18-0.24\mu$m; roughly corresponding to the {\it HST}/WFC3 F105W filter), and red colors show emission at $5.0-5.3\mu$m (rest-frame $0.99-1.05\mu$m; covering the blue part of the MIRI/F560W filter). The nuclear H$\alpha$+[N~II] emission overlaps with the nucleus detected by MIRI \citep{Colina23, CrespoGomez24}, while the diffuse rest-frame near-infrared emission extends until the outer H$\alpha$+[N~II] loop. We note that the Pa$\alpha$ emission detected by MIRI has a similar extent \citep{Bik23}. The rest-frame UV emission overlaps with the Western region detected in H$\alpha$+[N~II], including the location of the North-Western clumps.}
    \label{f:rgb}
\end{figure}

\begin{figure*}
    \centering
    \includegraphics[width=0.33\textwidth]{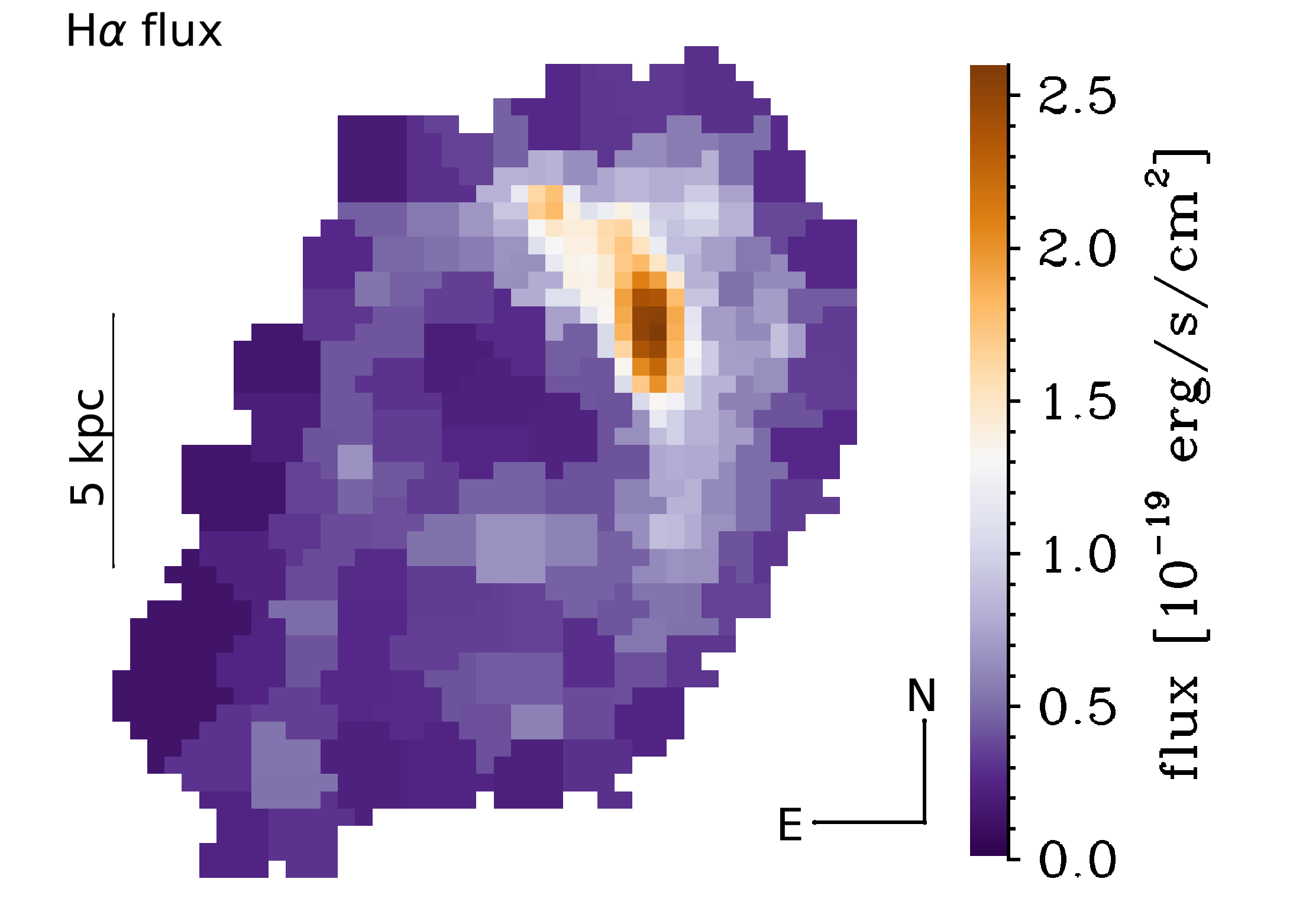}
    \includegraphics[width=0.33\textwidth]{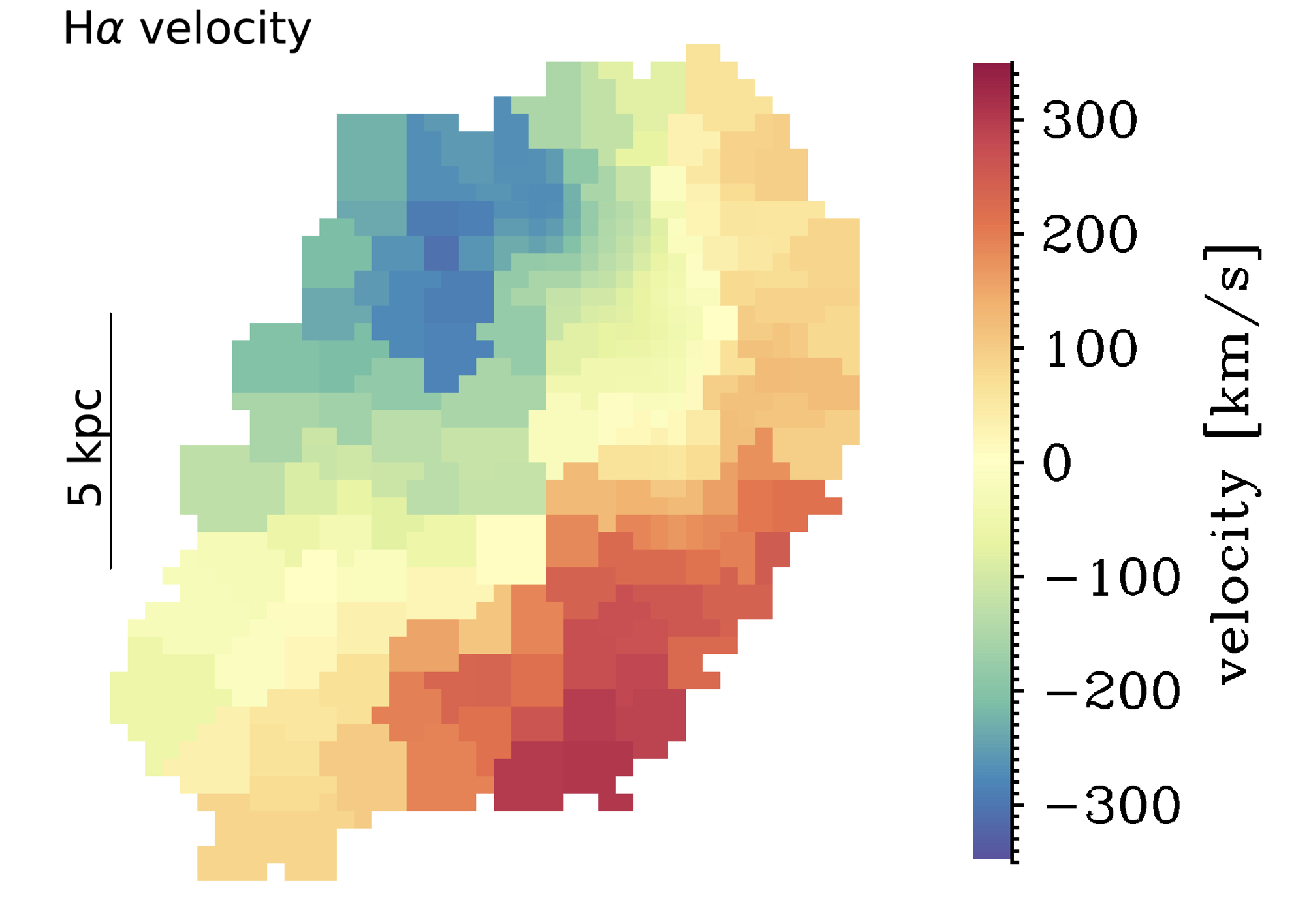}
    \includegraphics[width=0.33\textwidth]{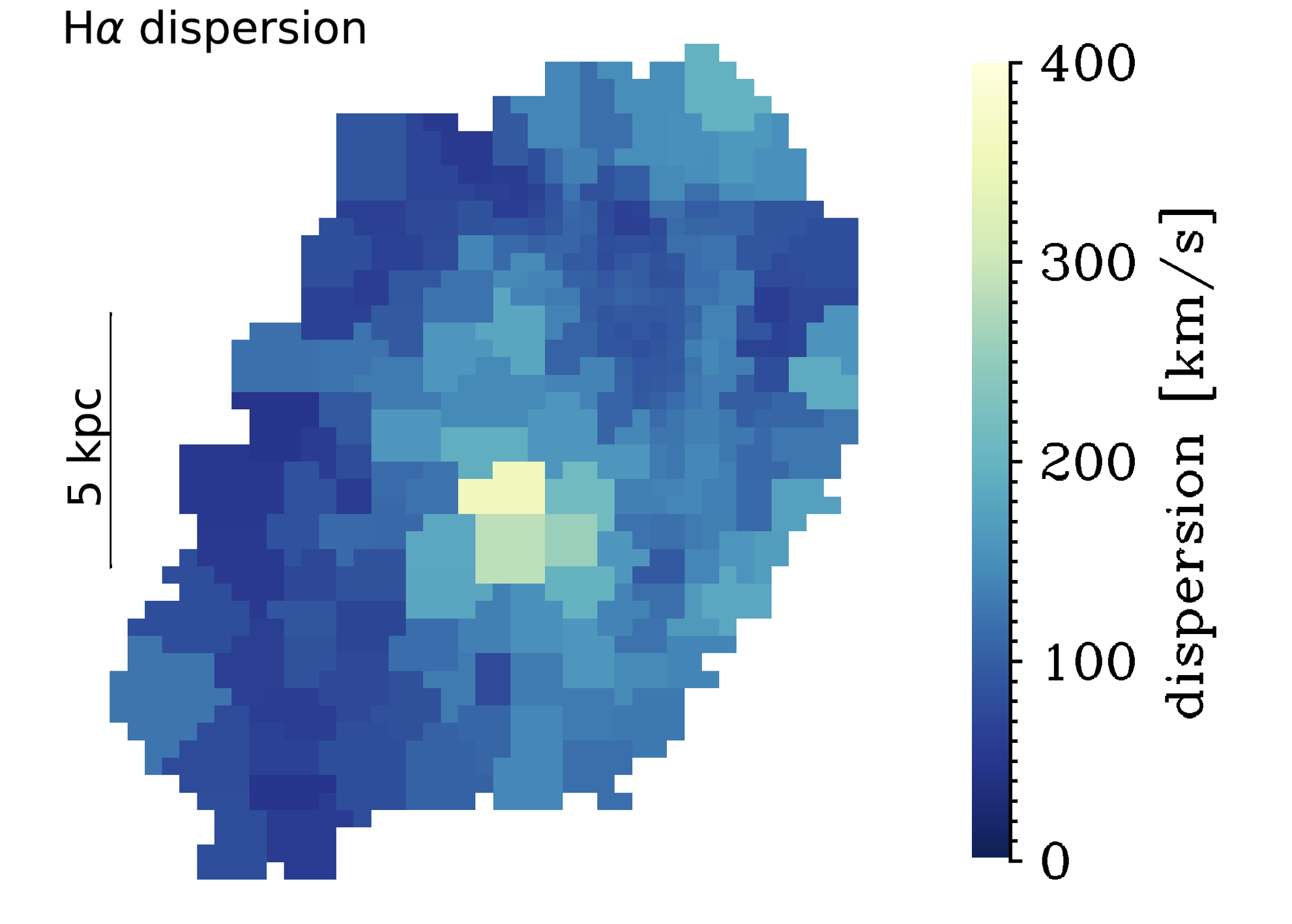}    \includegraphics[width=0.33\textwidth]{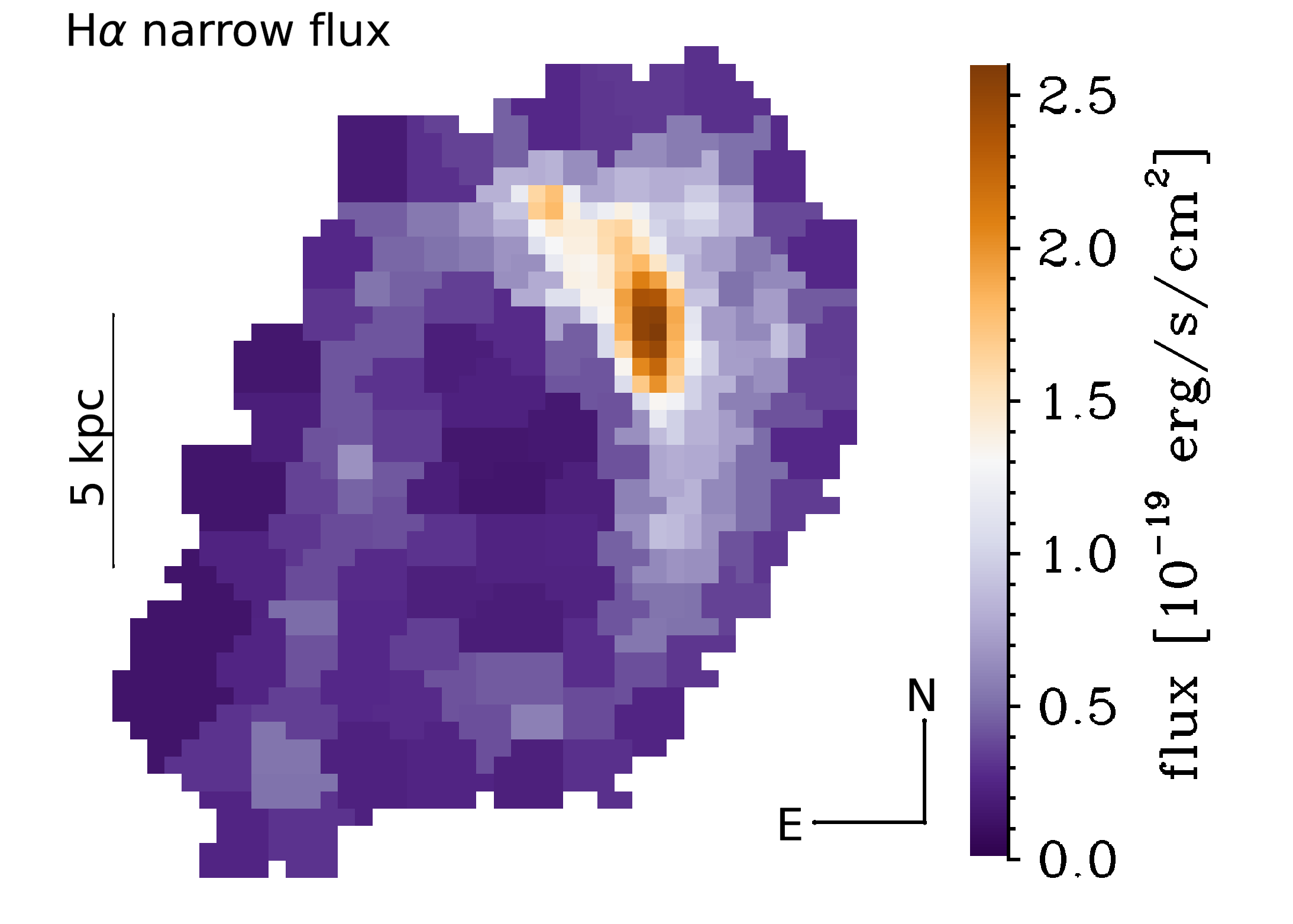}
    \includegraphics[width=0.33\textwidth]{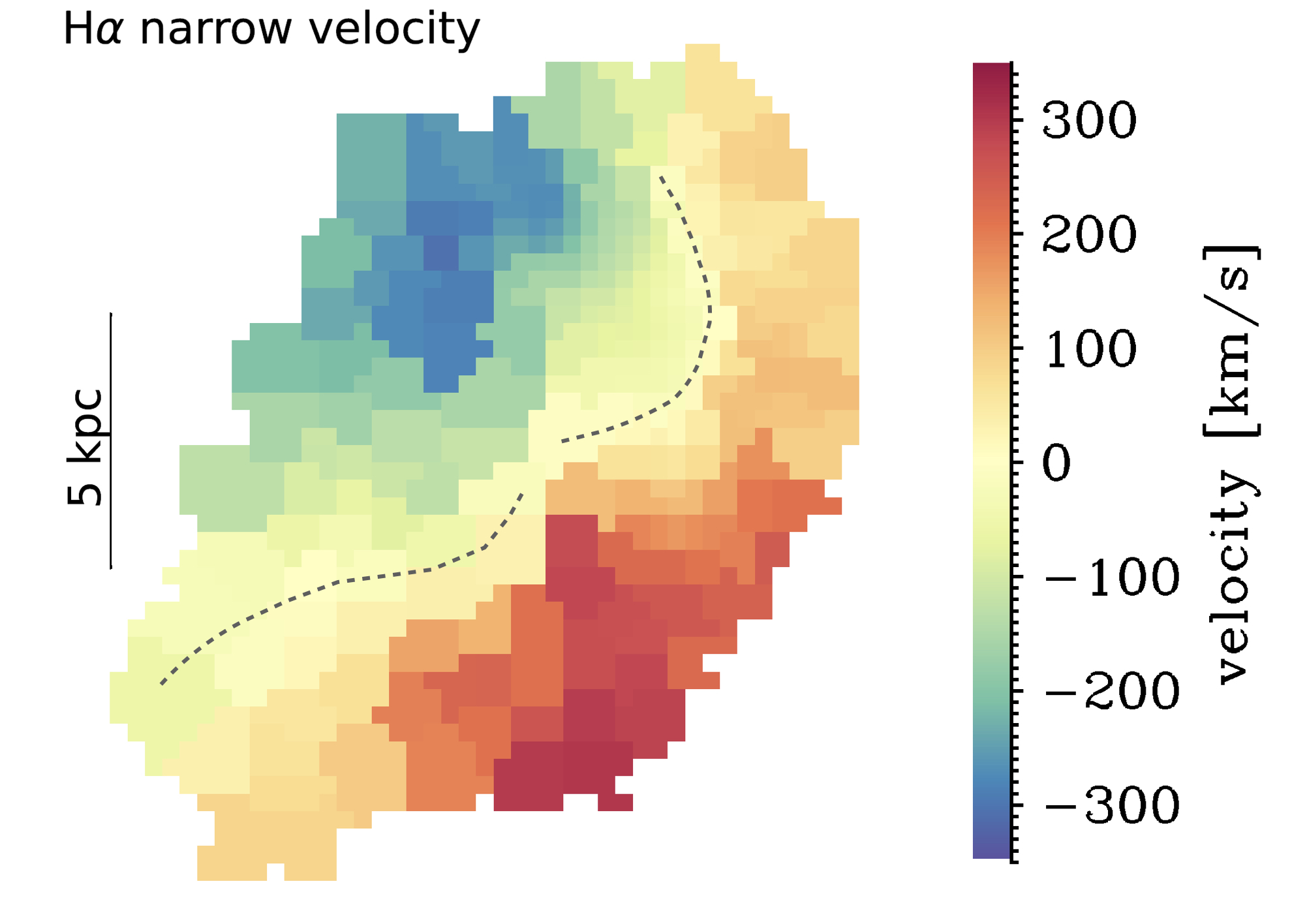}
    \includegraphics[width=0.33\textwidth]{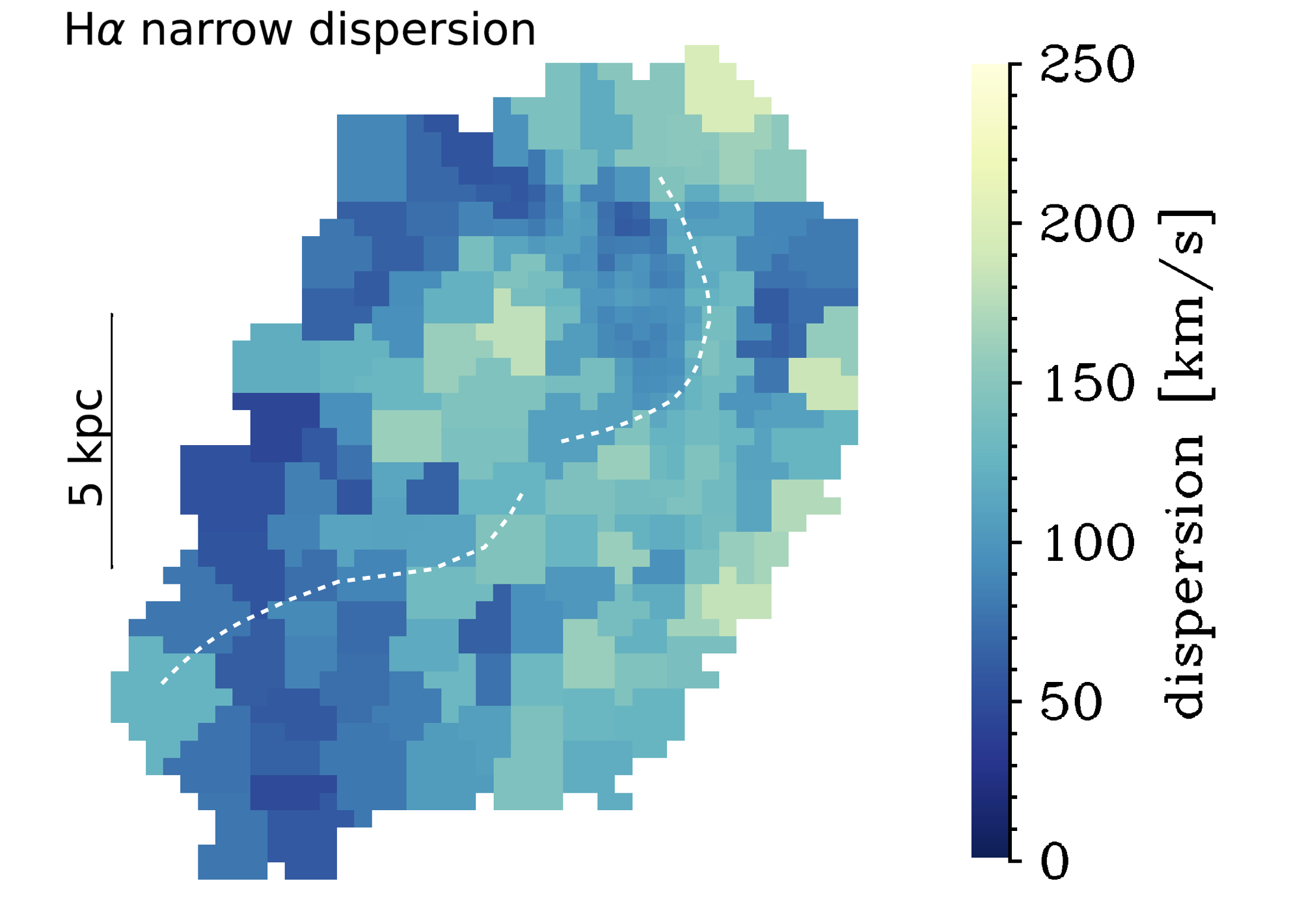}
    \caption{Top: Voronoi-binned maps (target $S/N\geq15$) of the observed H$\alpha$ flux (left), velocity (middle) and velocity dispersion (right) from a one-component fit. In the bottom row, we show corresponding maps where we have stitched the central region including only the narrow component of a two-component Gaussian fit (see Section~\ref{s:fitting}).
    The large-scale kinematics are consistent with a rotating disc. The high dispersion values in the central region in the top right panel trace a nuclear broad flux component. Removing this component, we find high average values of $\sigma_{\rm obs}\sim50-150$~km/s throughout GN20 (bottom right panel).
    We also find evidence for non-circular motions in the velocity field through patterns that deviate from the classical spider diagram expected for rotating discs: to guide the eye, we mark $v_{\rm obs}\sim0$~km/s as a dashed line in the bottom middle panel (repeated in the bottom right panel for reference).}
    \label{f:kinmaps}
\end{figure*}

\subsection{Dynamical modelling}\label{s:dyn}

To construct dynamical models, we use {\tt DysmalPy} (\citealp{Davies04a, Davies04b, Davies11, Cresci09, WuytsS16, Genzel17, Lang17, Uebler18, Price21, Genzel23}; L.~L.~Lee et al.~subm.), a 3D forward-modelling code that takes into account the instrumental effects of beam-smearing, line broadening, and finite spatial resolution. We build a mass model for GN20 informed through existing multi-wavelength constraints, in particular the recent MIRI observations at mid-infrared wavelengths that trace the stellar light distribution \citep{Colina23}. The model includes a baryonic disc and bulge component together with a spherical dark matter halo. For the baryonic mass distribution, we account for a finite flattening following \cite{Noordermeer08}. 

We fix the following structural parameters to the best-fit values derived by \cite{Colina23}: the disc effective radius $R_{e,\rm disc}=3.6$~kpc, the disc Sérsic index $n_{\rm S,disc}=0.42$, the minor-to-major axis ratio $b/a=0.80$. 
We adopt the convention that in the face-on case, the galaxy rotates in counter-clockwise direction for $i=0^\circ$, and in clockwise direction for $i=180^\circ$.
Assuming a thick disc with ratio of scale height to scale length of $q_0=0.2$ \citep[e.g.][]{WuytsS16, Genzel17}, $b/a=0.80$ corresponds to an inclination of either $38^\circ$ (projected counter-clockwise rotation; see also \citealp{CrespoGomez24}) or $142^\circ$ (projected clockwise rotation). 
Assuming that the regions in the North-West which are brightest in H$\alpha$ and UV emission are closer to us (the `near side'), this would suggest that GN20 rotates in a clockwise direction (see also stellar light distribution; \citealp{Colina23}). We therefore fix $i=142^\circ$.
We include a Gaussian prior on the total baryonic mass centred on $\log(M_{\rm bar, tot}/M_\odot)=11.4$, with a standard deviation of 0.3 and bounds of [10.4; 12.4]. This is motivated by estimates of the stellar mass \citep[$M_\star=1.1\times10^{11} M_\odot$;][]{Tan14} and of the molecular gas mass \citep[$M_{\rm H2}=1.3\times10^{11} M_\odot$;][here assuming $\alpha_{\rm CO}=0.8$]{Hodge12}. 
\cite{Colina23} identify an unresolved nuclear component (upper size limit 0.8~kpc). The authors conclude that this component corresponds to an obscured nuclear starburst, but they also discuss the alternative options of a massive stellar bulge or an AGN. To account in our model for this central mass component, we use their stellar mass estimate derived under the hypothesis of a bulge, $2.5\times10^{10} M_\odot$. This is about 10 per cent of our total baryonic mass estimate. We measure the effective radius of the nuclear component through a two-dimensional Gaussian fit to the H$\alpha$+[N~II] linemap, and find 0.76-0.80~kpc, consistent with the upper limit derived by \cite{Colina23} from the MIRI F560W imaging.
We model a round bulge with $R_{e,\rm bulge}=0.8$~kpc, $n_{\rm S,bulge}=4$ and a bulge-to-total baryonic mass ratio of $B/T=0.1$. 
We include a \cite{NFW96} dark matter halo with a concentration parameter of $c=3.1$ (the expected value for a halo of $\log(M_{\rm halo}/M_\odot)=13$ at this redshift, following \citealp{Dutton14}). We fit for the dark matter fraction within the effective radius, $f_{\rm DM}(r<R_e)$, from which the total halo mass is inferred \citep[see][]{Price21}. We use a flat prior $f_{\rm DM}(<R_e)=[0; 1]$. 
Further to the priors described above, we include a flat prior on the intrinsic velocity dispersion, $\sigma_0=[30; 150]$~km/s, which is assumed to be constant and isotropic throughout the disc.
In addition to the total baryonic mass $M_{\rm bar}$, the dark matter fraction at the effective radius $f_{\rm DM}(<R_e)$, and the intrinsic velocity dispersion $\sigma_0$, we allow the position angle to vary in the range PA=[$0^\circ$; $50^\circ$], and we allow the model to adjust the systemic velocity by $\pm100$~km/s.

We model the point-spread function (PSF) as a Gaussian with FWHM~$=0.12''$, corresponding to the approximate PSF (along slicers) at the wavelength of H$\alpha$ \citep[see][]{DEugenio23}. For the instrumental dispersion we adopt $\sigma_{\rm inst}=57$~km/s, which corresponds to the nominal spectral resolution at the wavelength of H$\alpha$ \citep{Jakobsen22}.

As input for our fiducial {\tt DysmalPy} runs, we use the Voronoi maps of the narrow H$\alpha$ velocity and velocity dispersion. 
We fix the centre of the model such that it falls onto the Voronoi bin with $v_{\rm obs}\sim0$~km/s along the axis connecting the observed velocity minimum and maximum. This coincides roughly with the centre of the disc component identified by \cite{Colina23}.
To account for the Voronoi binning during the minimization, we scale the uncertainties by a factor equal to the square root of the number of spaxels per bin.
We find the best fit through MCMC sampling, using 200 walkers and 500 steps after a burn-in of 200 steps \citep[][{\tt emcee}]{ForemanMackey13}. Our final chains are longer than 10 times the autocorrelation time for the individual fit parameters.
For visual comparison of the best-fit model to our data in the figures below, we show the median values of the model per Voronoi bin.

As motivated in Section~\ref{s:fuel}, we further construct a model with the same parameters as described above, but which additionally includes a uniform planar, radial inflow. The inflow velocity is not fitted for, but added as a fixed parameter \citep[see][]{Price21}. The preferred inflow velocity is initially identified through a grid search by injecting different values of $v_r$ in steps of 10~km/s from zero to 200~km/s, using least-squares minimization \citep{Markwardt09}. We repeat the fit with the so identified preferred inflow velocity ($v_r=130$~km/s) using MCMC. The results of this model are presented in Appendix~\ref{a:inflow}.

\section{Complex H$\alpha$ emission}\label{s:hamorph}

In Figure~\ref{f:halinemap} we show a line map of the emission integrated in the range $3.313-3.327\mu$m in our NIRSpec observations. This range covers the H$\alpha$+[N~II]$\lambda6584$ flux in GN20 (extended central object). 
Within the IFS field of view, we detect another galaxy North-West of GN20, which we call GN20b. It is at a projected separation of about 12~kpc from the centre of GN20 (also seen in the {\it HST} data; see Figure~\ref{f:rgb}) and a velocity difference of about +750~km/s, respectively a redshift of $z_{\rm H\alpha,GN20b}=4.06532\pm0.00003$. GN20b is also detected in continuum in the prism observations.

The map for GN20 reveals a complex, clumpy, loop-like structure. 
The emission is brightest in the North-Western region, close to the location of the rest-frame UV emission tracing young stars, previously detected with {\it HST} (see Figure~\ref{f:rgb}). 
Further to the North-West of this region, we detect two fainter clumps (dashed circles in Figure~\ref{f:halinemap}). In the remaining regions of GN20, the H$\alpha$ emission is much fainter. 
A central flux concentration largely overlaps with the nuclear component identified in the MIRI imaging \citep[][the black-edged white circle in their Figure~2]{Colina23}, which is also prominent in the NIRSpec data above $\sim5\mu$m (see Figure~\ref{f:rgb}). 
Immediately North of this region H$\alpha$ is barely detected. 
In the South-East, we detect clumpy H$\alpha$ emission extending to the South (dashed ellipse in Figure~\ref{f:halinemap}).

The peculiar surface brightness distribution of the H$\alpha$ emission in GN20 could spark doubts regarding its disc-like nature. However, we know from previous observations in the rest-frame far-infrared continuum that the central regions of GN20 are very dusty \citep{Hodge15}. This has also been shown through the recent analyses by \cite{Bik23, CrespoGomez24}, finding high attenuation in the central parts of the galaxy. 
In addition, the analysis of the stellar light distribution from MIRI imaging is consistent with a fairly smooth stellar disc plus a compact nuclear component \citep{Colina23, CrespoGomez24}.
Clumpy H$\alpha$ emission originating from smooth, disc-like mass distributions has also been observed at lower redshift \citep{WuytsS12}. Notably, most of these galaxies show smooth H$\alpha$ velocity fields consistent with disc rotation.
Indeed, this appears to be case as well for GN20: the H$\alpha$ kinematics, much like previous studies of GN20 utilising CO and Pa$\alpha$ as kinematic tracers, support the interpretation of a large rotating disc. We discuss the large-scale rotation and deviations from circular motions in Section~\ref{s:kinematics}.

The two H$\alpha$ clumps in the North-West appear also visible in CO(2-1) and in the UV \citep{Hodge15}, and there is also faint emission seen in the MIRI F560W image, on the level of a few per cent of the peak flux \citep[see][and Figure~\ref{f:rgb}]{Colina23}. 
Some F560W flux extends from the main disc along the same direction of the South-East extension \citep[see][]{Colina23}. Clumpy UV emission is also seen here, however this likely stems from a $z\sim1.74$ galaxy identified through rest-frame optical line emission in our prism data.
The H$\alpha$ clumps and extension could resemble real substructure embedded within GN20, e.g. young star-forming regions in the outer disc or a protruding spiral arm. Alternatively, they could correspond to fresh material in the process of accretion onto GN20. 
We note faint H$\alpha$ flux between GN20 and GN20b, which could further indicate a past or ongoing interaction.

\begin{figure*}
    \centering
    \includegraphics[width=0.245\textwidth]{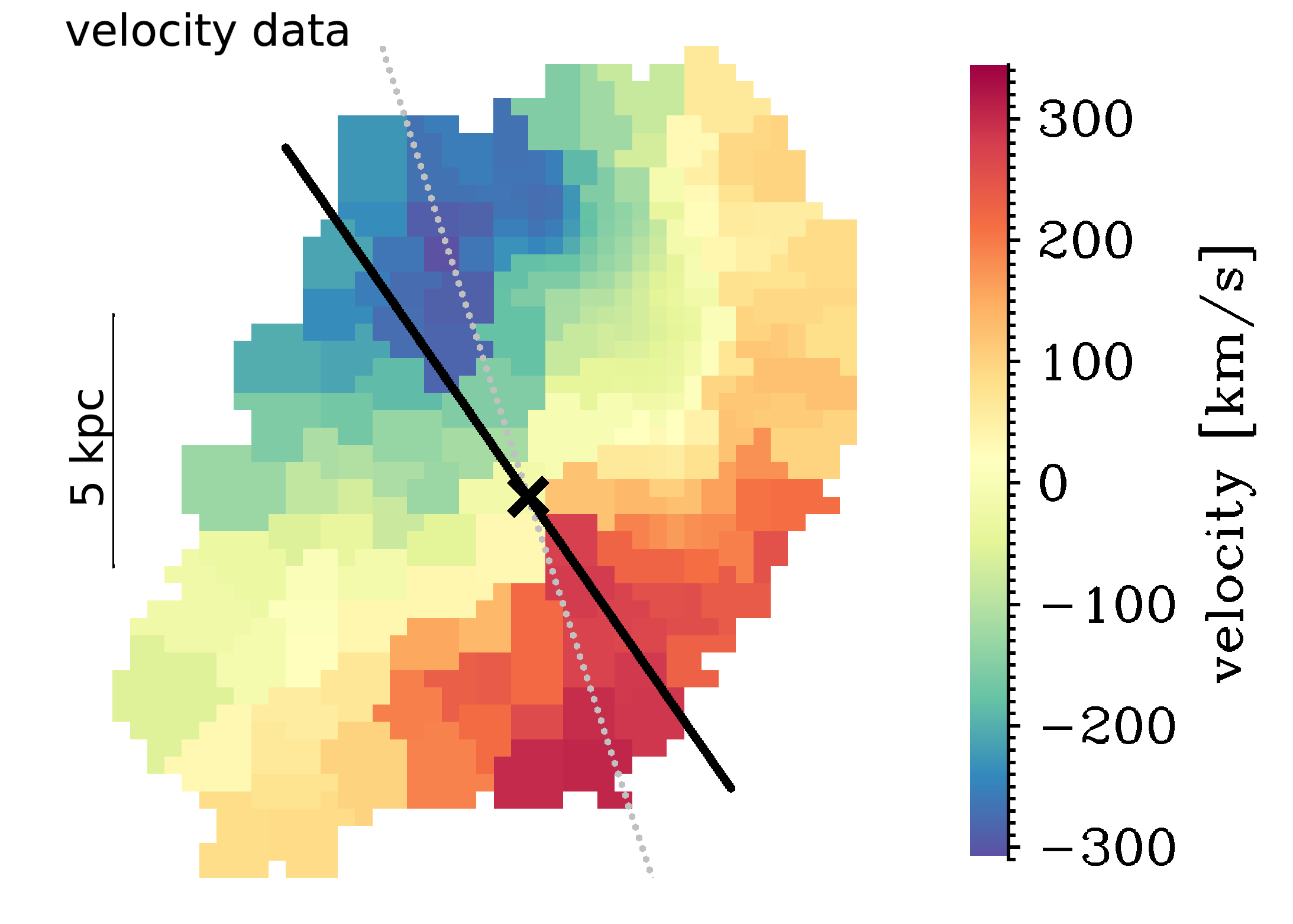}
    \includegraphics[width=0.245\textwidth]{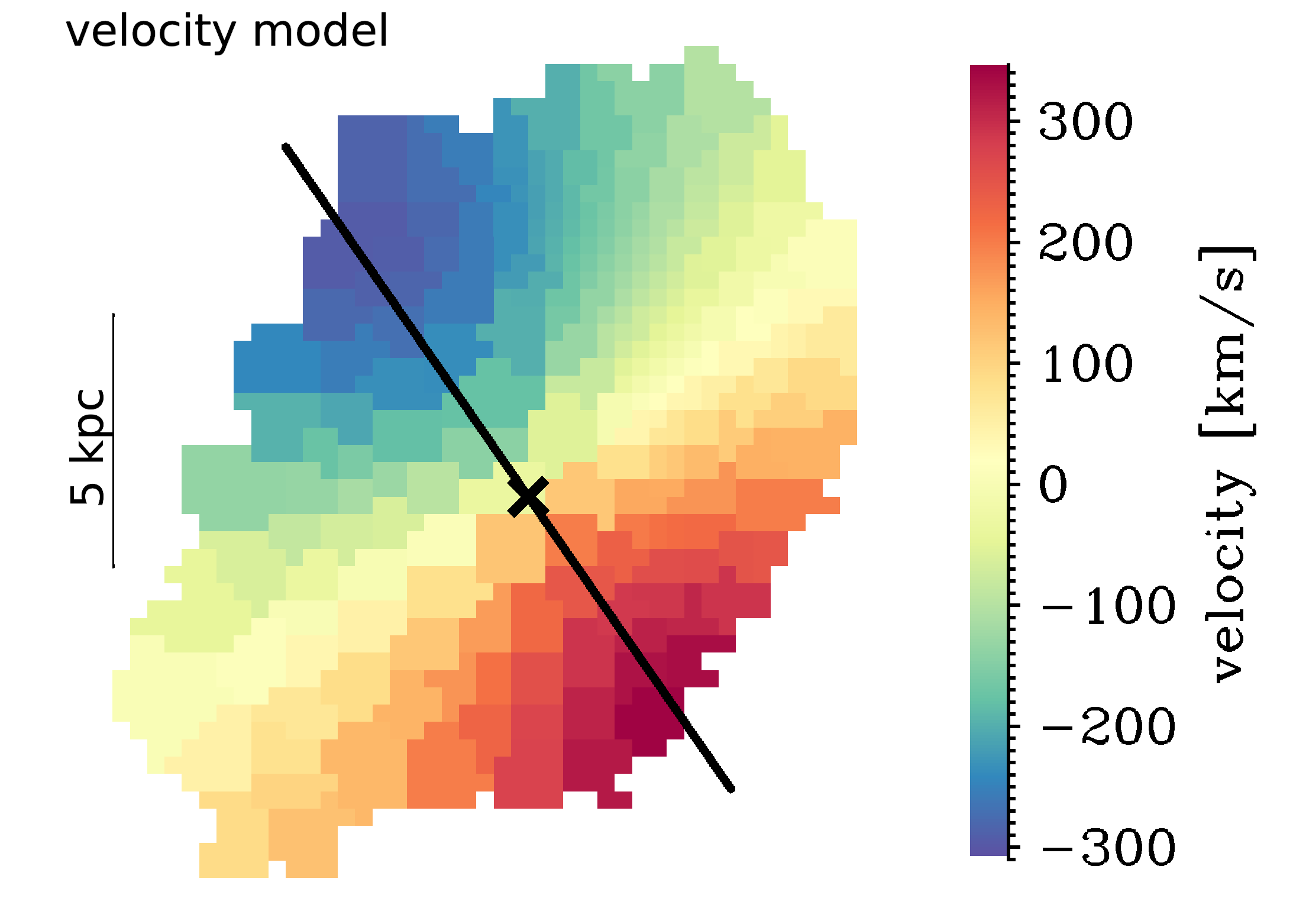}
    \includegraphics[width=0.245\textwidth]{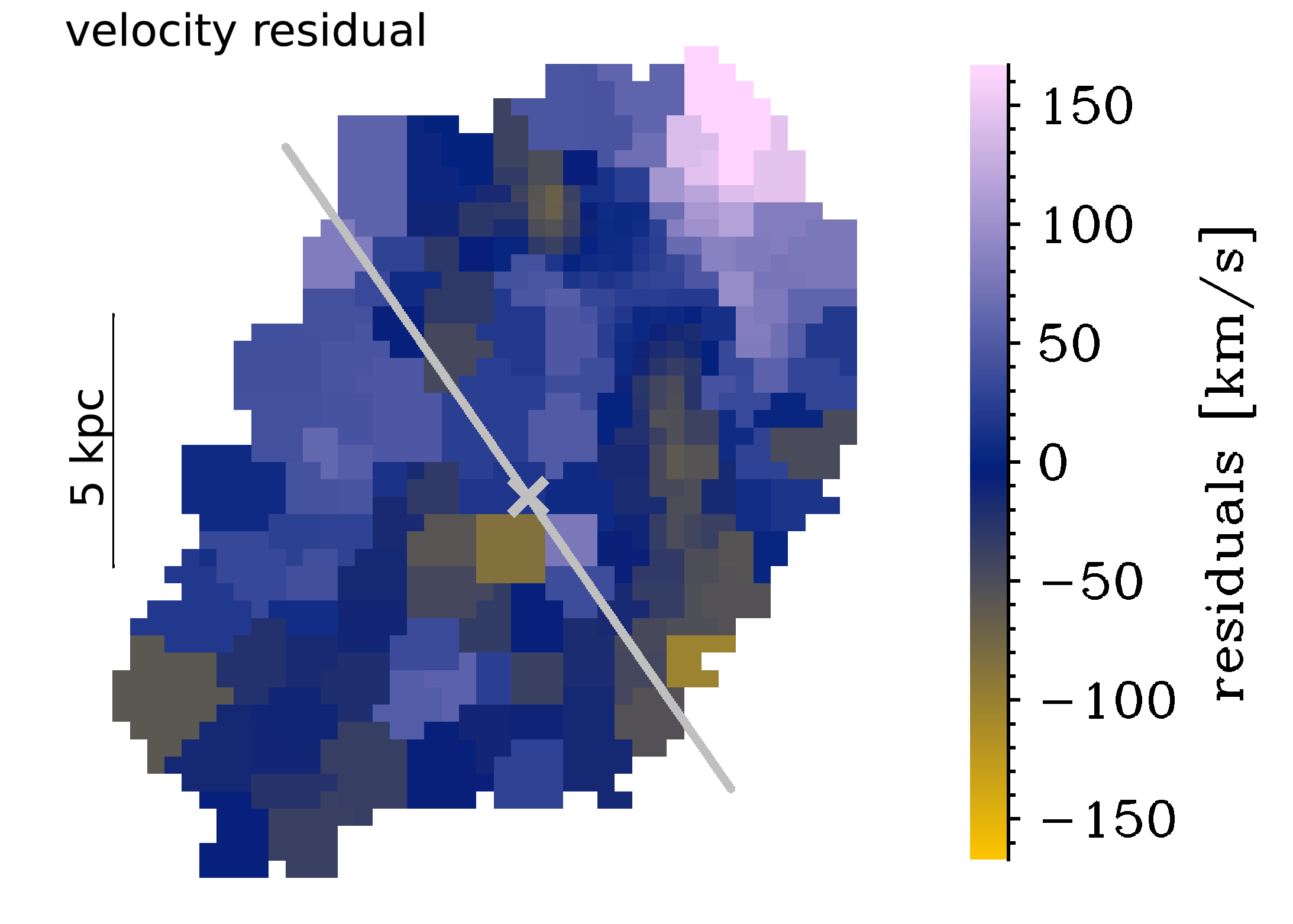}
    \includegraphics[width=0.245\textwidth]{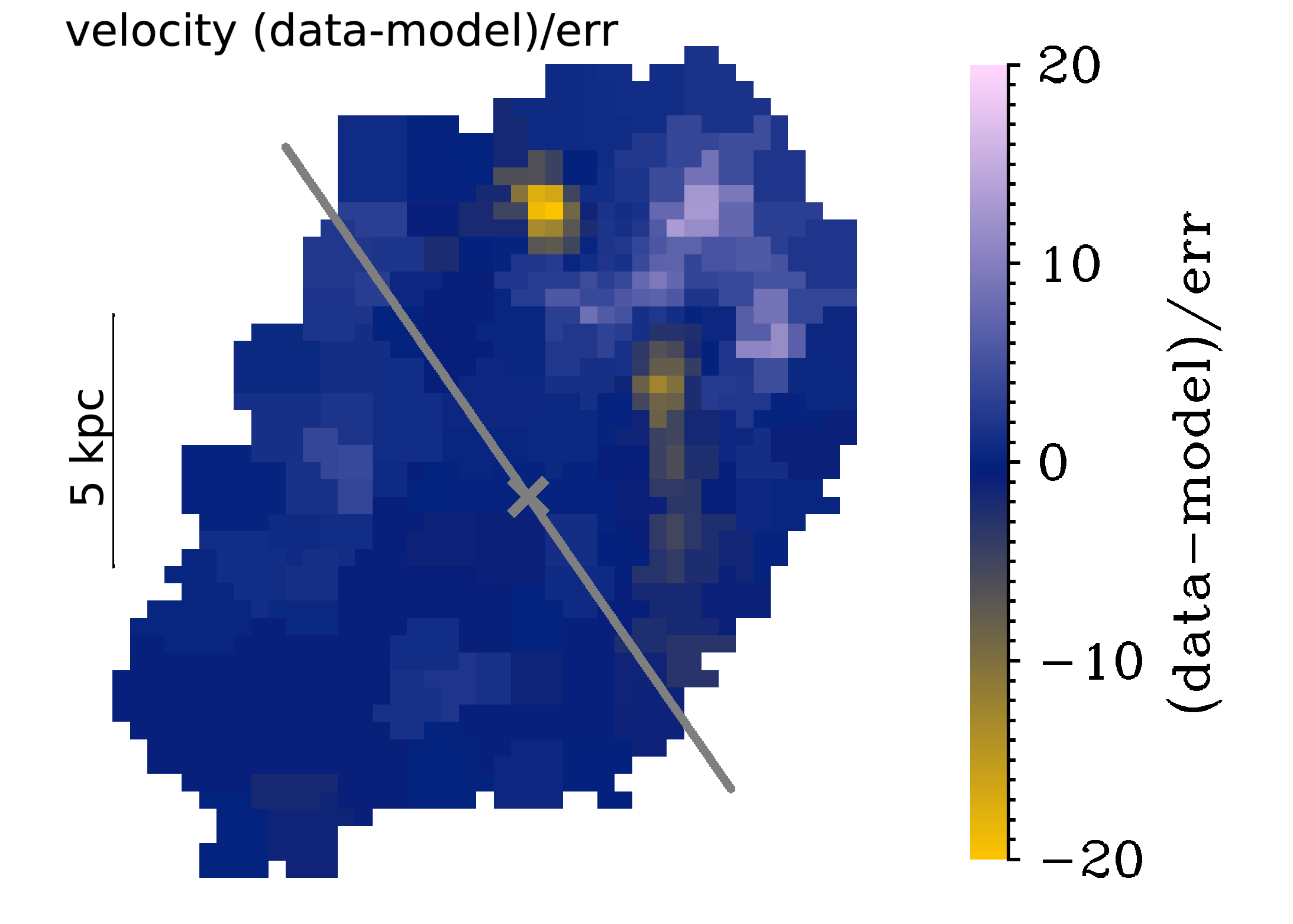}
    \includegraphics[width=0.245\textwidth]{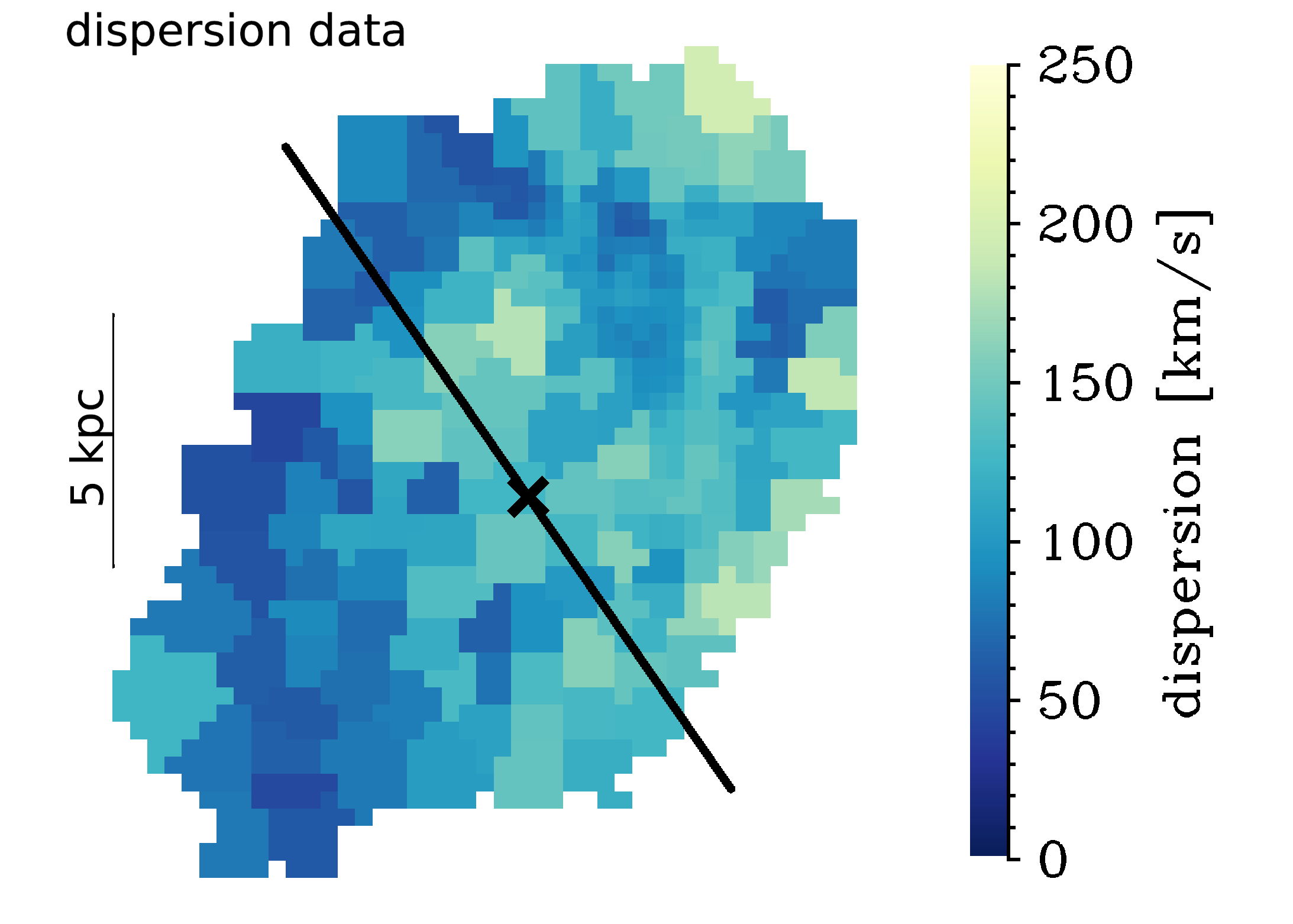}
    \includegraphics[width=0.245\textwidth]{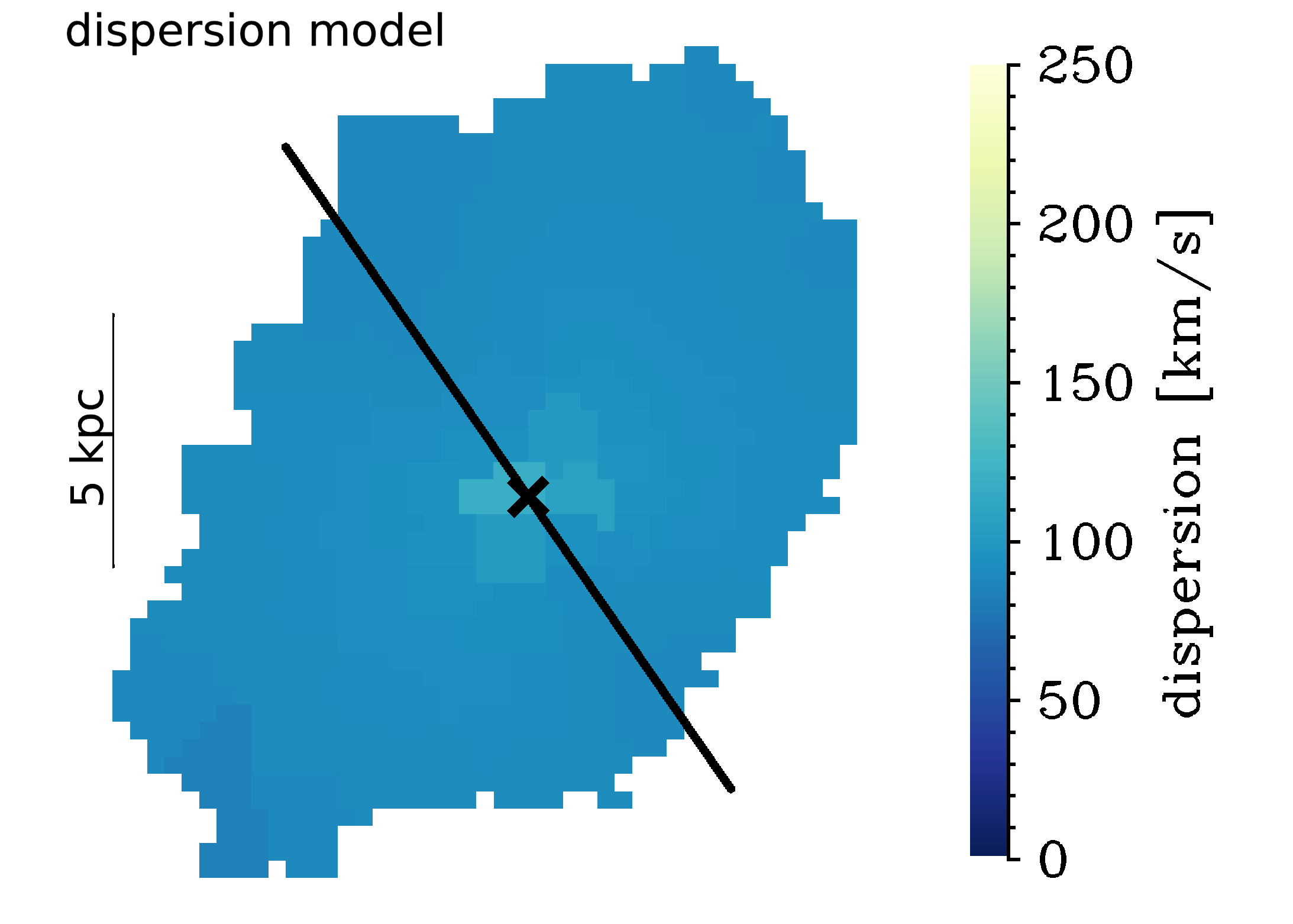}
    \includegraphics[width=0.245\textwidth]{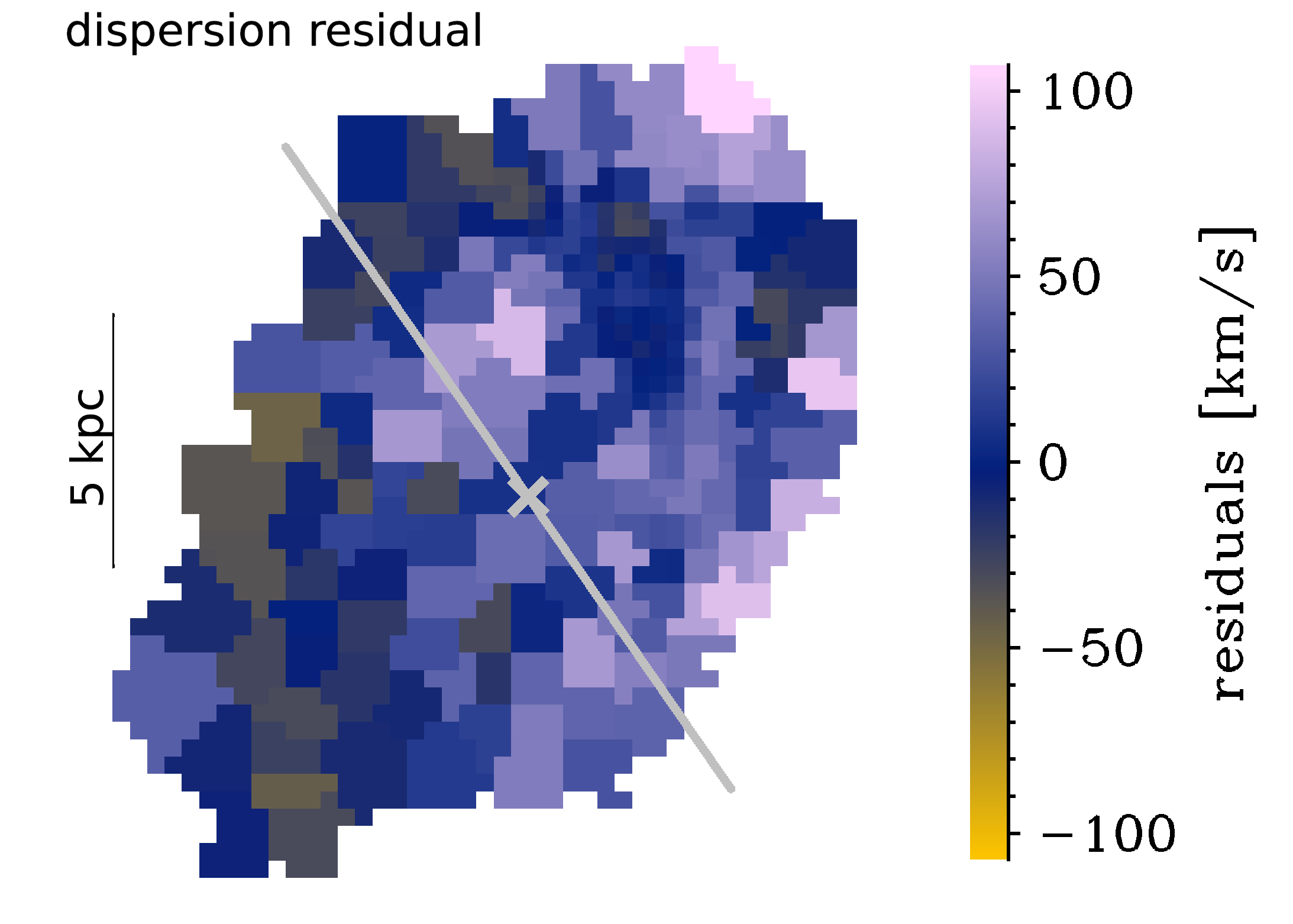}
    \includegraphics[width=0.245\textwidth]{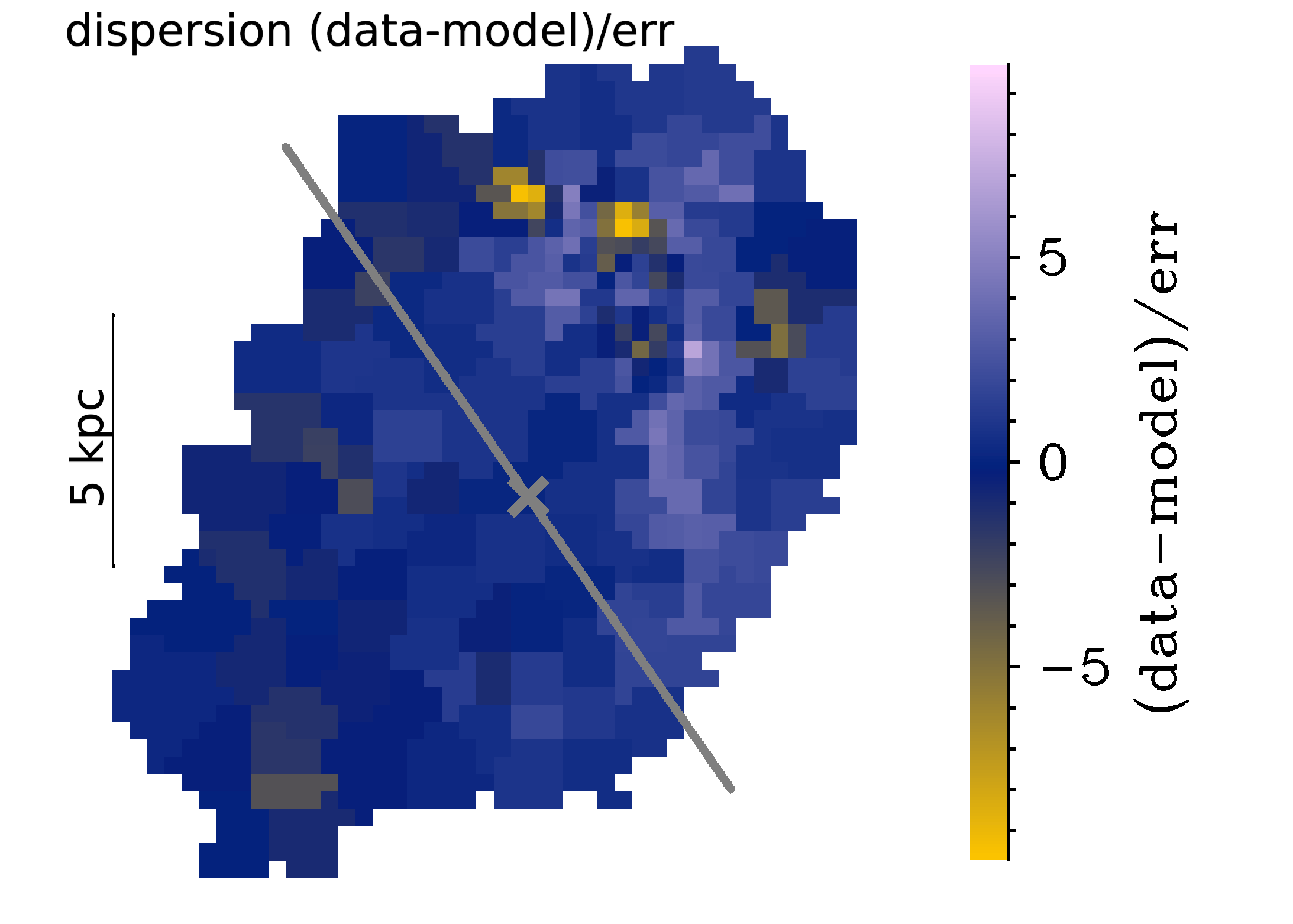}
    \caption{Our best-fit dynamical model including a disc, bulge and halo. Top (bottom) panels show the observed velocity (velocity dispersion) field (left), the best-fit model (middle left), residuals (data-model; middle right), and goodness-of-fit (data-model)/uncertainties (right). The cross and line indicate the centre and best-fit PA. For comparison, the dotted grey line in the top left panel indicates the axis connecting the observed velocity minimum and maximum. The model provides a reasonable fit to the data, with a median velocity offset of $\Delta v_{\rm med}=5.5$~km/s and a rms velocity difference of $\Delta v_{\rm rms}=49.3$~km/s. Corresponding values for the velocity dispersion are $\Delta\sigma_{\rm med}=14.9$~km/s and $\Delta\sigma_{\rm rms}=34.7$~km/s. Yet, we observe strong residuals particularly in the North-Western region, indicating a disturbance of the kinematics for instance through accretion or external perturbations.}
    \label{f:model}
\end{figure*}

\section{Large-scale rotation and non-circular motions}\label{s:kinematics}

In the middle and right panels of Figure~\ref{f:kinmaps} we show H$\alpha$ velocity and velocity dispersion maps based on our fits to the Voronoi-binned cube (top: one-component; bottom: stitched narrow component, see Section~\ref{s:fitting} for details).  
We find a large velocity gradient across GN20 of about $\Delta v_{\rm obs}\sim610$~km/s, and observed velocity dispersion values of $\sigma_{\rm obs}\sim50-150$~km/s in the Voronoi bins in the outer disc. The large-scale H$\alpha$ velocity field is consistent with disc rotation, and the kinematic major axis, observed velocities and velocity dispersions are in general agreement with earlier results based on molecular gas kinematics \citep{Carilli10, Hodge12}.
However, the data reveal clear deviations from circular motions in the velocity field. In the bottom middle panel of Figure~\ref{f:kinmaps}, we indicate $v_{\rm obs}\sim0$~km/s by dashed lines, illustrating a twist of the apparent kinematic minor axis.
Along the North-Western $v_{\rm obs}\sim0$~km/s twist we observe slightly elevated velocity dispersions, as evident from the bottom right panel of Figure~\ref{f:kinmaps}.

In the following, we further investigate the deviations from circular motions through residual analysis of a pure circular motion dynamical model. We also compare our modelling results to previous studies of the kinematics in GN20 based on other kinematic tracers.

\subsection{Dynamical modelling results and comparison to other kinematic tracers}\label{s:fiducialmodel}

We fit a dynamical model including a rotating disc, bulge, and dark matter halo, as described in Section~\ref{s:fitting}, to our data. The best-fit model and residuals are shown in Figure~\ref{f:model}. The model provides a reasonable fit to the data, with median offsets in velocity and velocity dispersion of $\Delta v_{\rm med}=5.5$~km/s and $\Delta\sigma_{\rm med}=14.9$~km/s. 
From this model, we measure a rotation velocity $v_{\rm rot}(R_e=3.6{\rm kpc})=469$~km/s (circular velocity $v_c(R_e)=496$~km/s), reaching a maximum value of $v_{\rm max}=531$~km/s ($v_{c,\rm max}=574$~km/s) at $r=6.2$~kpc ($r=6.8$~kpc). 
The circular velocity accounts for the effects of pressure support from turbulent motions, and is defined as $v_c^2(r)=v_{\rm rot}^2(r)+2\sigma_0^2 r/R_d$, with $R_d$ being the disc scale length \citep[see][]{Burkert10, Burkert16}.
We constrain an intrinsic velocity dispersion of $\sigma_0=89$~km/s. For a galaxy at $z\sim4.055$, from the relation by \cite{Uebler19}, calibrated based on ground-based observations up to $z\sim3.5$, we would expect an intrinsic ionised gas velocity dispersion of $\sigma_0\sim63\pm19$~km/s, somewhat lower than our best-fit result. However, we note that some regions in the outer disc of GN20 show comparable dispersion values in $\sigma_{\rm obs}$.
Evaluating the ratio of maximum rotation velocity to intrinsic velocity dispersion, we find $v_{\rm rot}(R_e)/\sigma_0=5.3$ ($v_{\rm rot,max}/\sigma_0=6.0$). 
The dynamical mass enclosed within one $R_e$ is $\log(M_{\rm dyn}(<R_e)/M_\odot)=11.3$, and within two $R_e$ is $\log(M_{\rm dyn}(<2R_e)/M_\odot)=11.7$. 
The inferred values from our best-fit model are reported in Table~\ref{t:model}.

To compare our modelling results to previous results on the kinematics of GN20 obtained in the literature, we repeat our dynamical modelling with the inclination fixed to the values adopted in the literature studies.
Assuming $i=45^\circ$, \cite{Carilli10} find a rotation velocity of $v_{\rm rot}(r\sim4{\rm kpc})=570$~km/s based on CO(6-5) molecular gas kinematics, with an enclosed dynamical mass of $M_{\rm dyn}(r<4{\rm kpc})=3\times10^{11}M_\odot$. 
Repeating our modelling with the inclination fixed to $i=135^\circ$ (i.e.\, corresponding to their $i=45^\circ$), we find a lower value of $v_{\rm rot}(r\sim4{\rm kpc})=442$~km/s, but a comparable enclosed total mass of $M_{\rm dyn}(r<4{\rm kpc})=2\times10^{11}M_\odot$.

\cite{Hodge12} and \cite{Bik23} adopt an inclination of $i=30^\circ$. From the CO(2-1) kinematics, \cite{Hodge12} inferred $v_{\rm max}=575\pm100$~km/s and $\sigma_0=100\pm30$~km/s. \cite{Bik23} found $v_{\rm max}(r<4~{\rm kpc})=550\pm40$~km/s and $\sigma_m=145\pm53$~km/s from Pa$\alpha$ kinematics, where $\sigma_m$ represents an upper limit on the intrinsic flux-weighted velocity dispersion.
Repeating our modelling with the inclination fixed to $i=150^\circ$ (i.e.\, corresponding to their $i=30^\circ$), we find a rotation velocity of $v_{\rm rot}(R_e=3.6{\rm kpc})=551$~km/s ($v_{\rm max}=646$~km/s), and $\sigma_0=89$~km/s. 
Our modelling results obtained by fixing $i=150^\circ$ are in good agreement with what has been inferred from these studies analysing molecular gas kinematics and Pa$\alpha$ kinematics on similar radial scales.
Comparing the amount of rotational support, \cite{Hodge12} find $v_{\rm max}/\sigma_0\mid_{\rm CO(2-1)}=5.8$, and \cite{Bik23} find a somewhat lower value of $v_{\rm max}/\sigma_m\mid_{\rm Pa\alpha}=3.8\pm1.4$, however based on the upper limit on the flux-weighted velocity dispersion. Using $i=150^\circ$, we find even higher values of $v_{\rm max}/\sigma_0=7.3$, yet at the effective radius (fixed to $R_e=3.6$~kpc), we find $v_{\rm rot}(R_e)/\sigma_0=6.2$. This is in general agreement with the literature results.
\cite{Hodge12} infer a dynamical mass of $M_{\rm dyn}=5.4\pm2.4\times10^{11}M_\odot$. Fixing $i=150^\circ$, we find $M_{\rm dyn}(r<R_e)=2.6\times10^{11}M_\odot$ and $M_{\rm dyn}(r<2R_e)=7.0\times10^{11}M_\odot$, in broad agreement with the results by \cite{Hodge12}.

Multi-phase gas kinematic measurements at high redshift are still rare. Recently, \cite{Parlanti23} measured $v_{\rm rot}(R_e=3{\rm kpc})/\sigma_0\mid_{\rm H\alpha}\sim9.2$ in an obscured AGN at $z=4.76$, with a higher value of $v_{\rm rot}(R_e)/\sigma_0\mid_{\rm [CII]]}\sim15.8$ measured from the [C~II] line \citep{Lelli21}. In this case, the authors speculate that the AGN might have deposited energy in the ionised gas phase, leading to higher dispersion values and therefore lower rotational support. 
In SPT0311-58, which is at the core of a massive proto-cluster, \cite{Arribas23} find $v_{\rm rot,max}/\sigma_0\mid_{\rm [OIII]}\sim1$ based on [O~III]$\lambda5007$ kinematics. Their median intrinsic velocity dispersion is $\sigma_0=113\pm19$~km/s, and in some regions 2-3 times higher than the velocity dispersion measured in [C~II]. This likely indicates that SPT0311-58 at $z=6.9$ is in an earlier phase of disc formation compared to GN20.

\begin{table}
\caption{Results from our best-fit dynamical model including a thick disc, bulge and dark matter halo. For $\sigma_0$, we adopt minimum uncertainties of 10~km/s. The effective radius is fixed to $R_e=3.6$~kpc.}
\begin{tabular}{lc}
\toprule
    \multicolumn{2}{c}{Fitted parameters}\\
    $\log(M_{\rm bar}/M_\odot)$ & $11.42^{+0.05}_{-0.06}$ \\
    $\sigma_0$ [km/s] & $89\pm10$ \\
    $f_{\rm DM}(<R_e)$ & $0.30\pm0.07$ \\
    PA$_{\rm kin}$ & $34.7^{+0.6}_{-0.3}$ \\
    velocity shift [km/s] & $29\pm2$ \\
    \hline
    $\chi^2_{\rm red}$ & 4.9728 \\    
    \hline
    \multicolumn{2}{c}{Derived values}\\
    $v_{\rm rot}(R_e)$ [km/s] & 469 \\
    $v_{\rm circ}(R_e)$ [km/s] & 496 \\
    $v_{\rm rot,max}$ [km/s] & 531 \\
    $v_{\rm rot}(R_e)/\sigma_0$ & 5.3 \\
    $\log(M_{\rm dyn}(<2R_e)/M_\odot)$ & 11.68 \\
\end{tabular} 
\footnotesize{}
\label{t:model}
\end{table}

\subsection{Deviations from circular motions}

Despite the relatively good fit of our circular motion model to the H$\alpha$ kinematics of GN20, and the agreement with literature results, we find strong residuals in particular in the North-Western region of the galaxy. 
This can be seen from the middle-right and right panels in Figure~\ref{f:model}.
These residuals are close to the H$\alpha$ clumps identified in Figure~\ref{f:halinemap}. Furthermore, the best-fit kinematic major axis appears tilted with respect to the velocity minimum and maximum in the data (see top left panel in Figure~\ref{f:model}). 
Deviations from circular motions in disc galaxies can be associated with a variety of phenomena: perturbations through past or ongoing interaction with neighbours, smooth accretion, streaming motions due to non-axisymmetric sub-structure like a bar or spiral arms, or outflows \citep[e.g.][]{vdKruit78, Roberts79, vAlbada81, Shlosman89, Athanassoula92, Wada92, Bournaud02, BT08, Tsukui24}.  

GN20 is known to be located within a proto-cluster environment together with the AGN GN20.2a and GN20.2b, located at projected distances of about $\sim150$~kpc and $\sim200$~kpc \citep{Daddi09}. 
Based on the projected separations and stellar masses of GN20.2a and GN20.2b, we do not expect a significant effect on the kinematics of GN20: we find a tidal strength parameter \citep{Dahari84, Verley07} of $Q=-3.9$, while sizeable effects are generally expected for $Q\geq-2$. 
However, as mentioned in Section~\ref{s:hamorph}, we detect another galaxy (GN20b) to the North-West of GN20 which could be involved in an ongoing or past interaction. 
It is therefore possible that deviations from circular motions in GN20 are triggered by past or ongoing interactions with neighbours.
This could also potentially explain the fairly large velocity dispersions found in some regions in GN20.
We note that offset nuclear emission detected in the MIRI imaging by \cite{Colina23} and \cite{CrespoGomez24} may also suggest a past gravitational interaction.

As mentioned in Section~\ref{s:fitting} and further discussed in Section~\ref{s:fuel}, we detect broad emission in the central region of GN20 which may be associated with high-velocity outflows. However, the impact of these non-circular motions is already removed in the kinematic maps used for the dynamical modelling (see Figure~\ref{f:kinmaps}).
Still, it is possible that past feedback activity in the centre of GN20, either by an AGN or an intense starburst, could have deposited energy in the interstellar medium, thus increasing the intrinsic velocity dispersion in the central parts of the galaxy \citep[see e.g.][]{Harrison16, Uebler19, Marasco23, Parlanti23}.
We do not expect any further impact on the narrow-component kinematics from outflows.

\begin{figure*}
    \centering
    \includegraphics[width=\columnwidth]{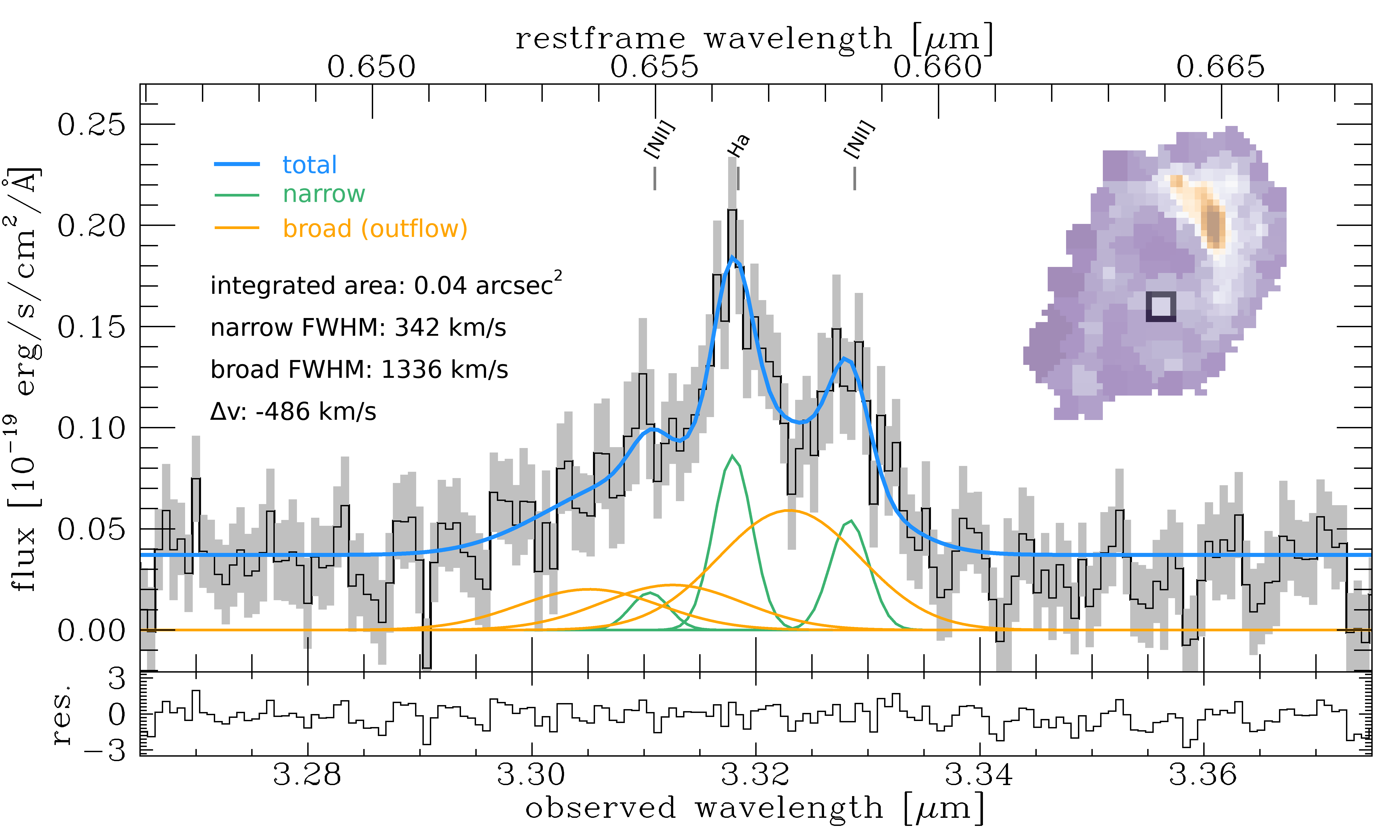}
    \includegraphics[width=\columnwidth]{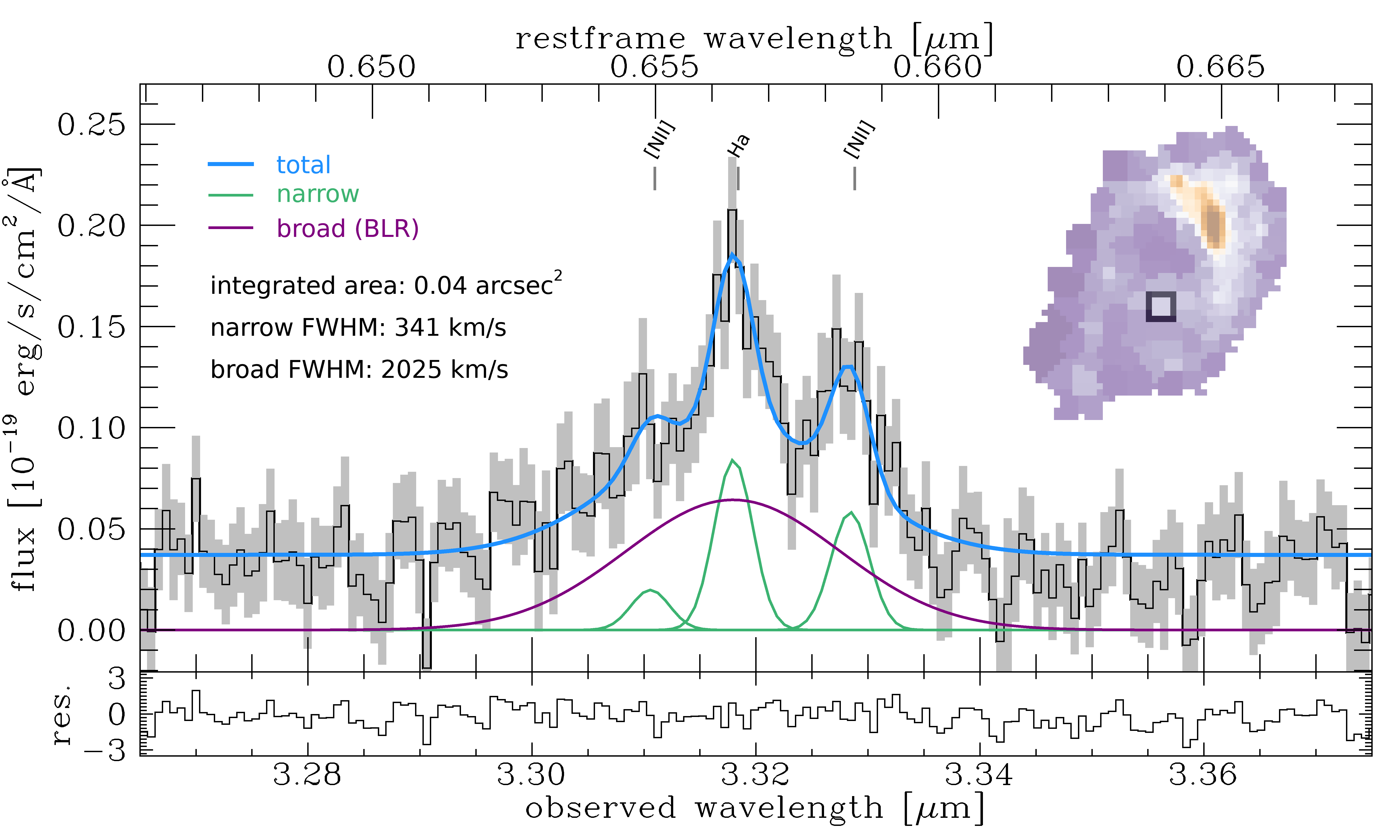}
    \includegraphics[width=\columnwidth]{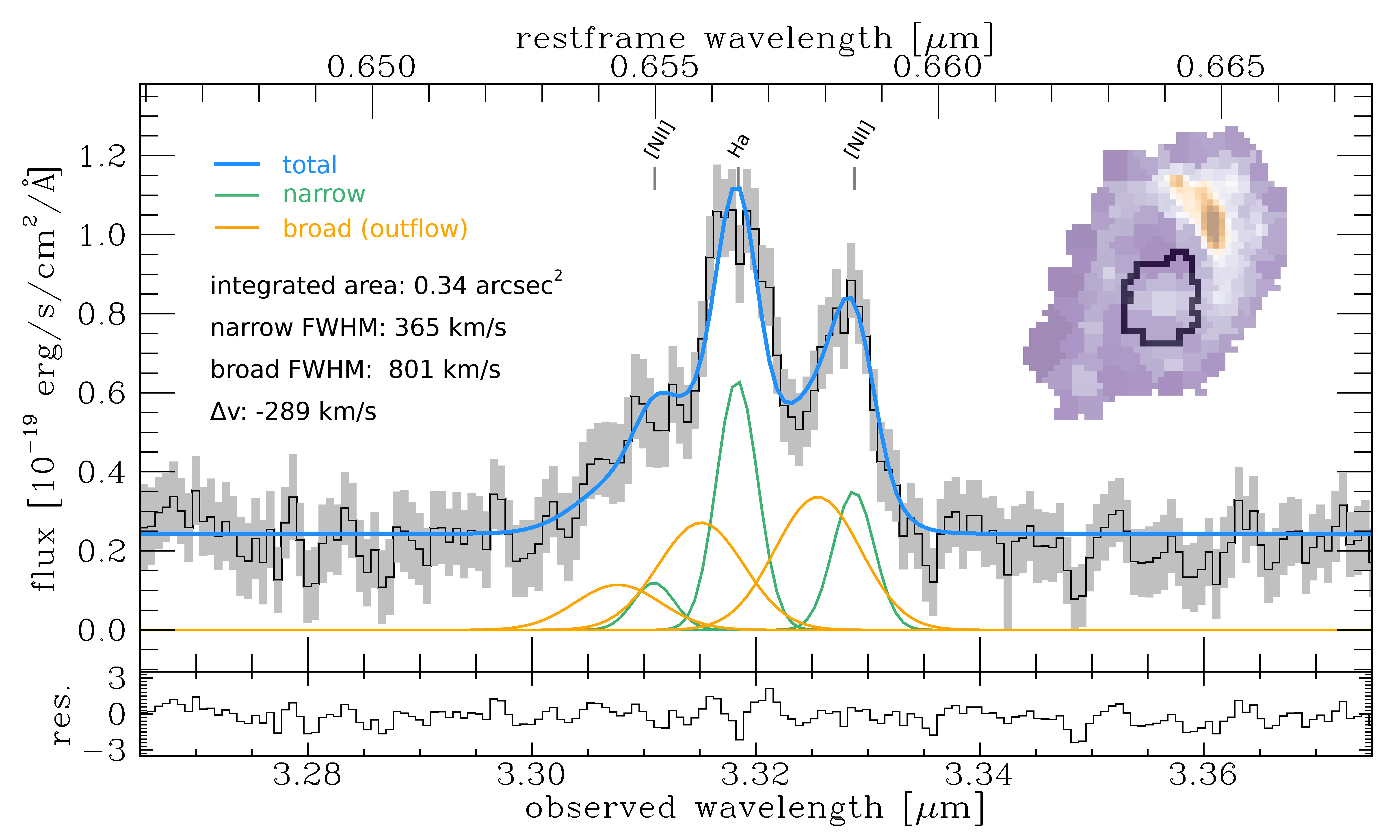}    \includegraphics[width=\columnwidth]{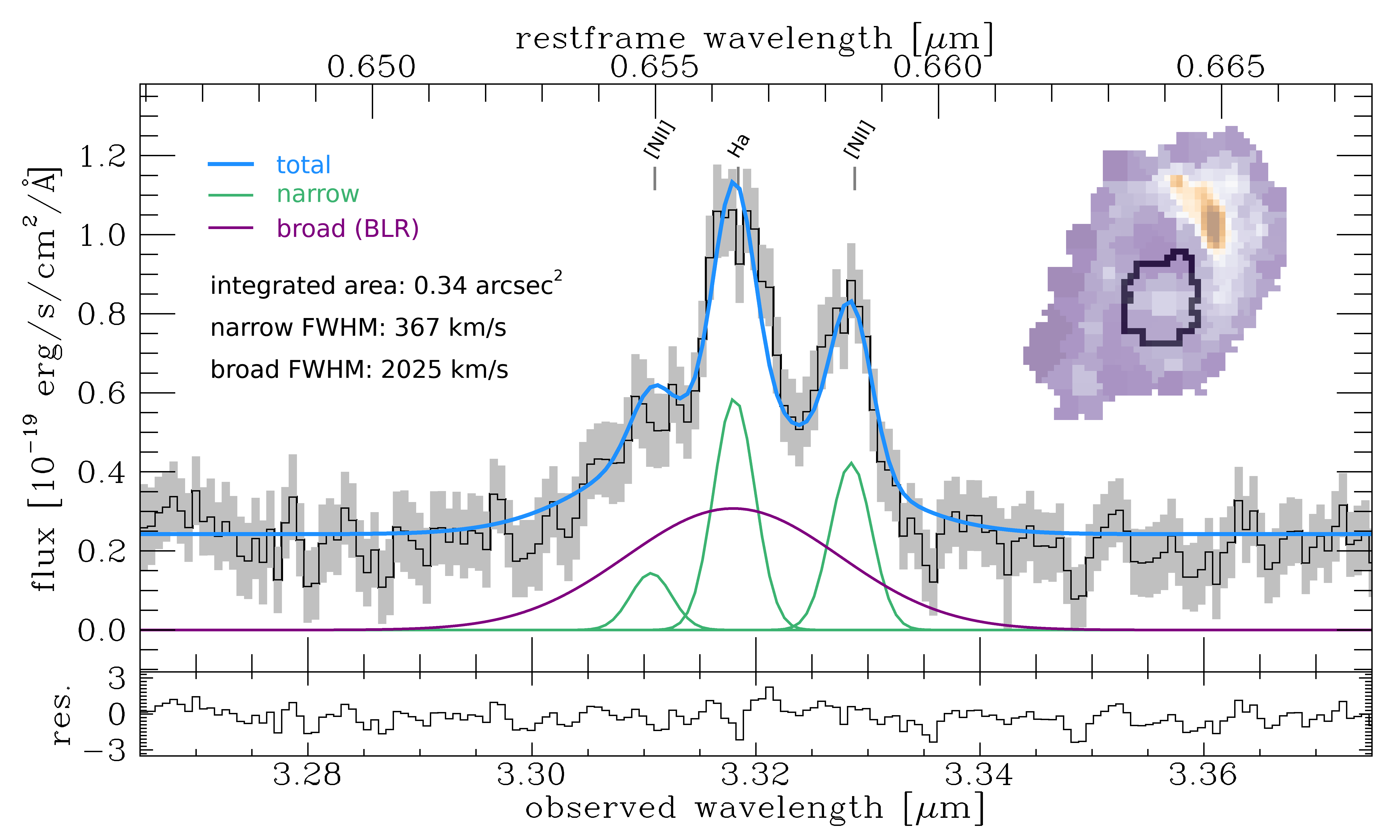}
    \caption{Two-component fits to integrated spectra in the central region of GN20. The top panels show fits to the central Voronoi bin with the brightest broad emission flux, while the bottom panel shows a fit to the larger aperture encompassing regions with broad emission based on visual inspection (see black contours in insets). 
    Left: fits including a set of broad components for H$\alpha$+[N~II]$\lambda\lambda$, i.e. interpreting the broad emission as an outflow. The broad emission is blue-shifted with respect to the narrow emission, with FWHM typical of AGN-driven winds, particularly in the central Voronoi bin (FWHM$_{\rm broad}>1000$~km/s). 
    Right: fits including a broad component only for H$\alpha$, i.e. interpreting the broad emission as the broad-line region of an accreting black hole. The two fitting setups have a comparable goodness-of-fit. This is also apparent from the very similar residuals shown in the bottom part of the panels, res.=(data-model)/uncertainties.
    The integrated spectra are not corrected for the velocity field.}
    \label{f:centralspec}
\end{figure*}

Finally, we consider non-circular motions in the form of radial flows. Those could be triggered through smooth accretion from the circumgalactic medium or internal streaming motions caused by disc instabilities or substructures such as spiral arms or a bar. The stellar light distribution derived through the MIRI F560W data by \cite{Colina23} does not provide evidence for a bar, though some substructure is revealed through their residual analysis which may indicate spiral-arm or ring-like features in the outer parts of the galaxy.
Based on our assumption of clockwise rotation (i.e. the bright NW side being the `near side', and the fainter SE side being the `far side'), the signature of an axisymmetric, planar radial inflow would be a mirrored S-shaped twist in the iso-velocity contours (see \citealp{vdKruit78}, and the recent examples at $z\sim2$ by \citealp{Genzel23} and \citealp{Price21}).
We observe a similar effect in the velocity field of GN20 (see bottom middle panel of Figure~\ref{f:kinmaps}). This motivates us to explore a second dynamical model including a uniform, planar radial inflow. This is clearly a simplified assumption for GN20, but it may still provide us with informative clues about the nature of the observed deviations from circular motions.
We show the results from this second model in Figure~\ref{f:inflow}. This model, which includes an inflow with $v_r=130$~km/s, indeed provides a somewhat better fit to the observed kinematics ($\Delta\chi^2_{\rm red}=0.33$), suggesting that gas may be flowing from the outskirts to the centre of GN20. However, the strong residuals in the NW largely remain, indicating that additional processes for instance recent or ongoing interaction disturb the kinematics in GN20.

\section{Broad nuclear emission and AGN signatures}\label{s:fuel}

As discussed in Sections~\ref{s:fitting}, we find broad emission in the H$\alpha$+[N~II] complex in the central region of GN20. To derive the narrow emission line maps we have used for the dynamical modelling, we have fitted this broad emission as a set of broad Gaussians for H$\alpha$ and [N~II]$\lambda\lambda6548,6583$ in addition to a set of narrow components, i.e.\, interpreting this broad emission as an outflow component. 
However, the data are very noisy due to the high obscuration in the centre. In fact, if we alternatively fit the central regions with only one broad component for H$\alpha$ in addition to the narrow line components for H$\alpha$+[N~II]$\lambda\lambda6548,6583$, we get spectral fits of similar quality. We, therefore, cannot exclude, given the quality of the data, that the broad emission in the centre traces the broad-line region (BLR) of an accreting black hole. Furthermore, more complex scenarios including both a potential BLR and an outflow component may be possible.
In the following we briefly discuss the two `extreme' scenarios of a pure outflow and a pure BLR, however we stress that we cannot robustly distinguish between these two scenarios based on our data.

\subsection{The outflow scenario}

Starting with the outflow interpretation, we show examples of our two-component fits to the Voronoi bin in the central region showing the brightest emission (top), and to the integrated spectrum extracted over a larger aperture where broad emission is visible (bottom) in the left panels of Figure~\ref{f:centralspec}.
In both regions (and indeed in all individual regions encompassed by the larger aperture) the broad components are blue-shifted. Their FWHM of 820-1340~km/s are typical of AGN-driven winds \citep[e.g.][and references therein]{Veilleux05, Veilleux20, Fabian12, Heckman14, Genzel14b, Harrison16, Carniani15, Rupke17, FS19}. From the large integrated aperture, we calculate a maximum outflow velocity of $v_{\rm out}=\langle v_{\rm broad}\rangle+2\sigma_{\rm broad}=970\pm220$~km/s, where $\sigma_{\rm broad}$ is corrected for instrumental resolution \citep[e.g.][]{Genzel11, DaviesRL19}. The large outflow velocities indicate that the outflow is driven by an AGN \citep[see also discussion by][]{Maiolino23a}.

Assuming a photo-ionised, constant-velocity spherical outflow \citep{Genzel11, Newman12, FS19, DaviesRL19, DaviesRL20, Cresci23} of the extent $R_{\rm out}$ equal to the aperture radius ($\sim2$~kpc), with an electron density of $n_{e, \rm out}=1000$/cm$^{3}$ \citep[e.g.][]{Perna17, Kakkad18, FS19}, we would find a low mass-outflow rate of $\dot{M}_{\rm out,ion}=0.7\pm0.3~M_\odot$/yr. We note that this estimate is uncertain due to the unknown outflow geometry and electron density in the outflow (the data in the central region is too noisy for a direct measurement from the [S~II] doublet).
Certainly, this value would correspond to a lower limit, considering the high obscuration in the centre of GN20, and because we are only tracing the warm ionised gas phase \citep[e.g.][]{Rupke13, HerreraCamus19, RobertsBorsani20, Fluetsch21, Avery22, Baron22, Cresci23, Belli23, Davies24}. For comparison, in a recent study of another obscured AGN at $z=4.76$, \cite{Parlanti23} find a mass outflow rate of $\dot{M}_{\rm out,ion}=11^{+57}_{-5.5}~M_\odot$/yr.

\subsection{The BLR scenario}

In the right panels of Figure~\ref{f:centralspec}, we show the corresponding fits including only one broad component fixed at the position of the narrow H$\alpha$ emission in the central Voronoi bin (see upper panel). The fit to the larger aperture is shown in the bottom, where we keep the position and FWHM of the broad component fixed to the fitting results from the central Voronoi bin.
The fits are of comparable quality to the fits interpreting the broad emission as an outflow ($\Delta$BIC$_{\rm outflow-BLR, central}=14$, $\Delta$BIC$_{\rm outflow-BLR, large~aperture}=9$).
From the properties of the (unobscured) putative BLR component, we would derive a black hole mass of $\log(M_\bullet/M_\odot)=7.3\pm0.4$ following \cite{Reines15} and using the fit to the large aperture.
For a galaxy of the stellar mass of GN20 ($M_\star\sim1.1\times10^{11} M_\odot$), a value of $\log(M_\bullet/M_\odot)\sim7.3$ would fall well within the scatter of local BLR AGN. However, this estimate does not account for extinction towards the BLR, and may therefore correspond to a lower limit, $\log(M_\bullet/M_\odot)\gtrsim7.3$.

Assuming that GN20 hosts an obscured AGN, \cite{Riechers14} estimated an Eddington limit for its black hole mass of $\log(M_\bullet^{\rm Edd}/M_\odot)=8.1-8.5$ based on its $6\mu$~m continuum luminosity and an upper limit on the $2-10$~keV $X-$ray luminosity. This value is consistent with our lower limit.

\begin{figure}
    \centering
    \includegraphics[width=0.495\columnwidth]{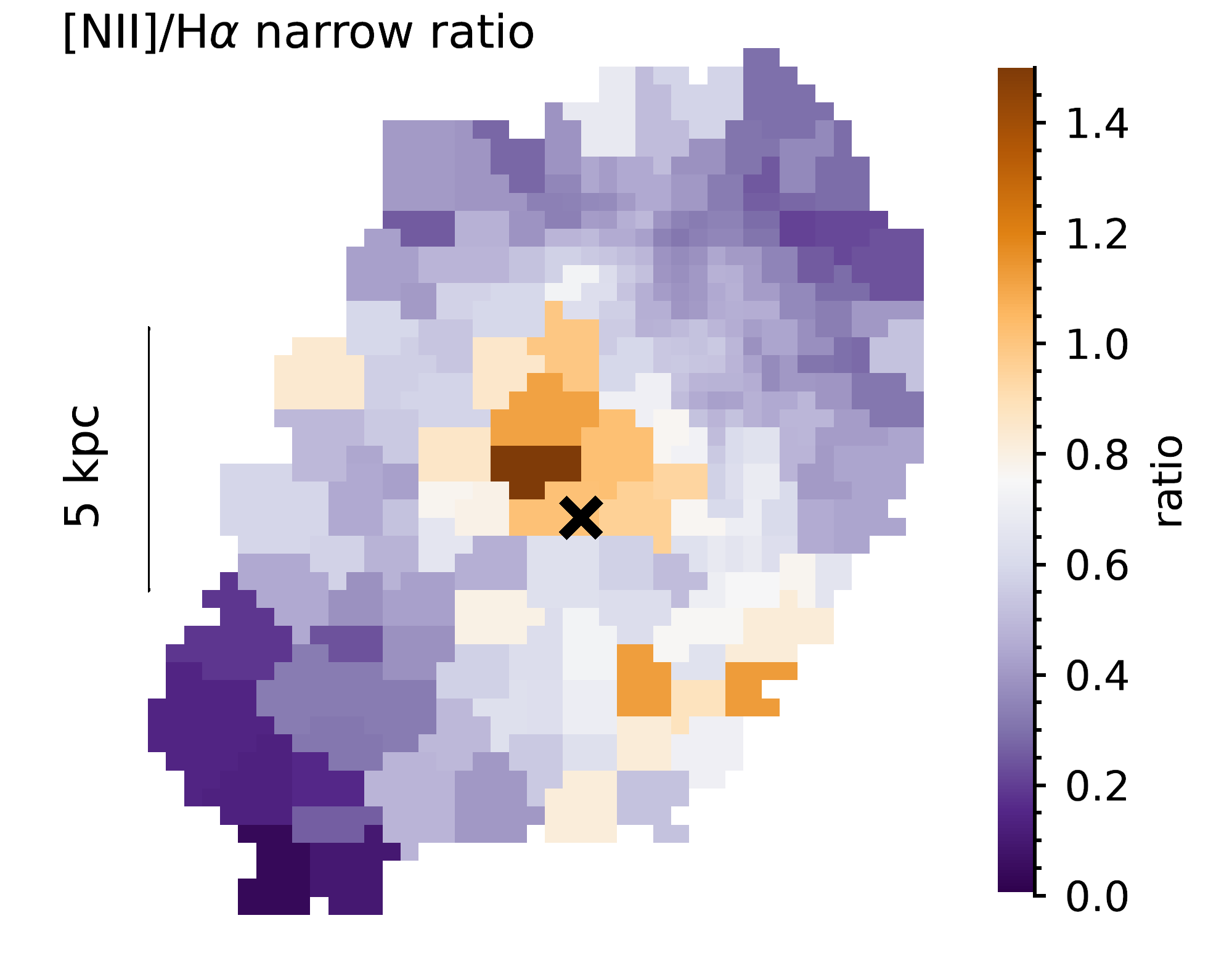}
    \includegraphics[width=0.495\columnwidth]{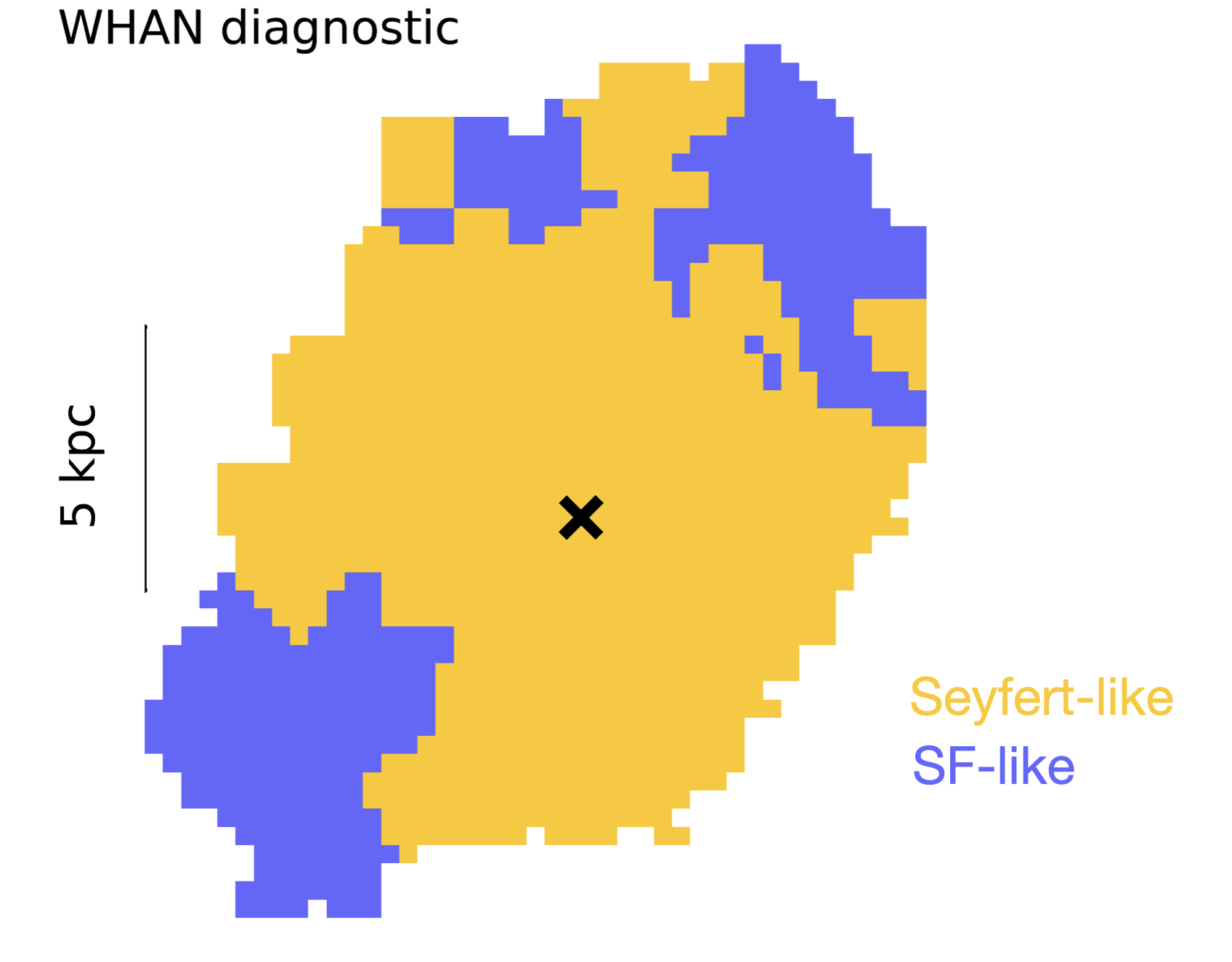}
    \caption{Left: map of the narrow-component [N~II]$\lambda6583$/H$\alpha$ ratio derived from fits to the H$\alpha$ Voronoi-binned map (see Section~\ref{s:fitting}).
    High ratios of [N~II]$\lambda6583$/H$\alpha>0.6$ throughout large parts of GN20, and especially in the Northern central regions and towards the South-West, indicate likely photo-ionisation by an AGN. 
    Right: GN20 as classified through the WHAN diagnostic diagram (see Section~\ref{s:diagnostics}). GN20 is largely consistent with the Seyfert regime, while some regions in the outskirts are consistent with being photo-ionised by star formation (SF).
    The black cross indicates the galaxy centre as adopted for the dynamical modelling.}
    \label{f:niiha}
\end{figure}

\subsection{Emission line diagnostics}\label{s:diagnostics}

In addition to the broad linewidths in the central region of GN20, suggesting either large outflow velocities or the presence of an actively accreting black hole, we find high narrow-component ratios of [N~II]$\lambda6583$/H$\alpha>0.6$ throughout large parts of GN20, and especially in the Northern nuclear region and toward the South-West. This is illustrated in the left panel of Figure~\ref{f:niiha}, where we show a map of the narrow-component [N~II]$\lambda6583$/H$\alpha$ ratio based on the Voronoi-binned map (see Section~\ref{s:fitting}).
Values of [N~II]$\lambda6583$/H$\alpha>0.6$ may be indicative of contributions through shocks, $\alpha-$enhanced evolved stellar populations, or photo-ionisation by an AGN \citep{Baldwin81, Kewley01, Kauffmann03, Byler19}. 

The BPT diagram \citep{Baldwin81} comparing [N~II]$\lambda6583$/H$\alpha$ to [O~III]$\lambda5007$/H$\beta$ can provide insights into the dominant ionisation mechanism in galaxies. The [O~III]$\lambda5007$ and H$\beta$ lines are not covered in our grating observations. They are covered in the prism observations, yet due to the high obscuration they are undetected in most regions \citep[see also][]{Maseda24}. As an alternative, we use the WHAN diagnostic diagram, which utilises the equivalent width of H$\alpha$ in combination with the [N~II]$\lambda6583$/H$\alpha$ ratio \citep{CidFernandes10, CidFernandes11}. In this diagram, for EW(H$\alpha)>6$~\AA\, and log([N~II]$\lambda6583$/H$\alpha)>-0.4$, galaxies fall into the Seyfert regime. Indeed, we find that most regions in GN20 are consistent with the Seyfert regime, while some regions in the outskirts are classified as star-forming (SF; EW(H$\alpha)>3$~\AA\ and log([N~II]$\lambda6583$/H$\alpha)<-0.4$). We show the classification based on the WHAN diagram throughout GN20 in the right panel of Figure~\ref{f:niiha}.

\section{Summary and Discussion}\label{s:conclusions}

We have presented an analysis of the rest-frame optical kinematics in the large $z\sim4.055$ sub-mm galaxy GN20 using {\it JWST}/NIRSpec-IFU data. In the following, we discuss and summarise our main findings.

{\bf H$\alpha$ morphology.} We find extended, but clumpy and asymmetric H$\alpha$ emission, that is largely arranged in a loop around a central emission peak. H$\alpha$ is brightest along the loop in the North-Western region (see Figure~\ref{f:halinemap}). The central H$\alpha$ emission peak broadly coincides with the compact rest-frame $1.1\mu$m nuclear (stellar) emission found through MIRI/F560W imaging by \cite{Colina23} \citep[see also][]{CrespoGomez24}. The edge of the extended envelope identified in their analysis overlaps with the H$\alpha$ loop. The brightest H$\alpha$ emission region partly overlaps with the rest-frame UV emission obtained with {\it HST} in F105W, but is generally located further to the North-East. We compare the rest-frame optical line emission obtained through our NIRSpec-IFS observations with rest-frame UV and near-infrared wavebands in Figure~\ref{f:rgb}.

{\bf Large-scale H$\alpha$ kinematics.} We find smooth, large-scale rotation in H$\alpha$ ($\Delta v_{\rm obs}\sim610$~km/s; $\sigma_{\rm obs,disc}\sim50-150$~km/s; Figure~\ref{f:kinmaps}). We construct a three-dimensional dynamical model including a thick disc, bulge, and dark matter halo, and simultaneously model the observed velocity and velocity dispersion maps by minimization on the projected two-dimensional kinematics. From our best-fit model (Figure~\ref{f:model}), we find $v_{\rm rot}(R_e)=469$~km/s, $\sigma_0=89$~km/s, $v_{\rm rot}(R_e)/\sigma_0=5.3$, and $\log(M_{\rm dyn}(<2R_e)/M_\odot)\sim11.7$. These results are largely consistent with previous analyses of the molecular and ionised gas kinematics constrained by VLA observations of CO(2-1), PdBI observations of CO(6-5), and MIRI/MRS observations of Pa$\alpha$ \citep{Carilli10, Hodge12, Bik23}. 

{\bf Non-circular motions.} The H$\alpha$ velocity field shows deviations from circular motions. These deviations are particularly apparent in the North-Western region of GN20, as also confirmed through a residual analysis based on our best-fit dynamical model (Figure~\ref{f:model}). Non-circular motions may arise from a variety of phenomena, and several may be acting simultaneously in GN20, potentially including tidal interactions, accretion and radial motions. To test one scenario, we construct a second dynamical model including a planar radial inflow of $v_r=130$~km/s, which indeed provides a slightly better fit to the data ($\Delta\chi^2_{\rm red}=0.33$). However, significant residuals in the North-West remain also with this model, indicating that additional causes for perturbation of simple rotation exist in GN20  (Figure~\ref{f:inflow}).

{\bf AGN signatures.} We find broad (FWHM$\sim1000-2000$~km/s) emission in the H$\alpha$+[N~II] complex in the central region of GN20. This broad emission can be modelled as a blue-shifted, high-velocity outflow in H$\alpha$ and [N~II] ($v_{\rm out}=970\pm220$~km/s), or as a broad-line region in H$\alpha$ emission, corresponding to a black hole of mass $\log(M_\bullet/M_\odot)\gtrsim7.3$ (Figure~\ref{f:centralspec}). We cannot robustly distinguish between these scenarios (or a combination of both) based on the data quality, however, both scenarios provide evidence for the presence of an active black hole in GN20. 
Further evidence is provided through high values of narrow-component [N~II]$\lambda6583$/H$\alpha>0.4$ together with EW(H$\alpha)>6$~\AA\, throughout large regions of the galaxy, consistent with photo-ionisation by an AGN (Figure~\ref{f:niiha}).

{\bf Evidence for black hole feeding and feedback.}
Our kinematic and dynamical analysis provides tentative evidence that gas may be channeled into the nuclear region of GN20. These inflows could potentially fuel central star formation and/or accretion onto the central black hole. Indeed we find evidence for an AGN in the centre of GN20 in the form of broad nuclear emission, signaling either AGN-driven outflows or an actively accreting black hole, and Seyfert-like ionisation throughout the galaxy.
In the local Universe, evidence for AGN feeding through molecular gas inflows has been observed, for instance through gravity torques induced by spiral or bar structure \citep[e.g.][]{Maiolino00, Hunt08, Casasola11, Combes14, Combes19, Speights16, Venturi18, Audibert19, Izumi23}.
At higher redshift, $1<z<2.5$, signatures of both ionised and molecular gas inflows have been revealed in deep observations of massive main-sequence galaxies \citep{Price21, Genzel23}.

In summary, our analysis suggests that GN20, an intriguingly large and massive SMG 1.5 billion years after the Big Bang, shows evidence for both non-circular motions (potentially fuelling AGN activity), and high ionisation and possibly high-velocity outflows (i.e. feedback by the AGN), or the broad-line region of an actively accreting black hole.
This work together with the recent studies by, for instance, \cite{Arribas23, Jones23, Parlanti23} highlights the potential of {\it JWST}/NIRSpec-IFS to unveil detailed, spatially-resolved properties of $z>4$ dusty star-forming galaxies, such as the source(s) powering nebular emission and high-accuracy kinematic measurements, even in cases of extremely high dust attenuation. While SMGs were previously mainly characterised through their bright sub-mm emission, {\it JWST} opened up new avenues to study their formation and evolution. As argued by several authors, the rapid evolution, high ongoing star formation, and dense environment make GN20 a prime progenitor candidate for present-day massive early-type galaxies, those systems that contain the majority of the stellar mass existing today.
Deeper NIRSpec observations of GN20 could help to further uncover the source of its broad nuclear emission, and to investigate in greater detail the deviations from circular motions in the H$\alpha$ kinematics. This might shed further light on the assembly and evolution history of today's most massive galaxies.

\section*{Acknowledgements}

We are grateful to the anonymous referee for a constructive report that helped to improve the quality of this manuscript.
We thank Chris Carilli and Luis Colina, Arjan Bik and the MIRI GTO team for sharing data.
We thank Arjan Bik and Takafumi Tsukui for comments on an earlier version of this manuscript.
H\"U thanks the Infrared/Submillimeter Group at MPE for making the latest version of {\tt DysmalPy} available prior to the code becoming public. Special thanks go to Stavros Pastras and Juan Manuel Espejo Salcedo for assistance with the code installation.
AJB and GCJ acknowledge funding from the ``FirstGalaxies'' Advanced Grant from the European Research Council (ERC) under the European Union’s Horizon 2020 research and innovation program (Grant agreement No. 789056).
BRdP, MP and SA acknowledge grant PID2021-127718NB-I00 funded by the Spanish Ministry of Science and Innovation/State Agency of Research (MICIN/AEI/ 10.13039/501100011033).
FDE, JS and RM acknowledge support by the Science and Technology Facilities Council (STFC), by the ERC through Advanced Grant 695671 ``QUENCH'', and by the UKRI Frontier Research grant RISEandFALL.
GV and SCa acknowledge support from the European Union (ERC, WINGS,101040227).
H{\"U} gratefully acknowledges support by the Isaac Newton Trust and by the Kavli Foundation through a Newton-Kavli Junior Fellowship.
IL acknowledges support from PID2022-140483NB-C22 funded by AEI 10.13039/501100011033 and BDC 20221289 funded by MCIN by the Recovery, Transformation and Resilience Plan from the Spanish State, and by NextGenerationEU from the European Union through the Recovery and Resilience Facility.
%

%%%%%%%%%%%%%%%%%%%%%%%%%%%%%%%%%%%%%%%%%%%%%%%%%%
\section*{Data Availability}
The NIRSpec data used in this research has been obtained within the NIRSpec-IFU GTO programme GA-NIFS and is publicly available at MAST.
Data presented in this paper will be shared upon reasonable request to the corresponding author.



\bibliographystyle{mnras}
\bibliography{literature} % if your bibtex file is called example.bib

%%%%%%%%%%%%%%%%%%%%%%%%%%%%%%%%%%%%%%%%%%%%%%%%%%

%%%%%%%%%%%%%%%%% APPENDICES %%%%%%%%%%%%%%%%%%%%%

\appendix

\section{Dynamical model including a fixed radial inflow}\label{a:inflow}

\begin{figure*}
    \centering
    \includegraphics[width=0.24\textwidth]{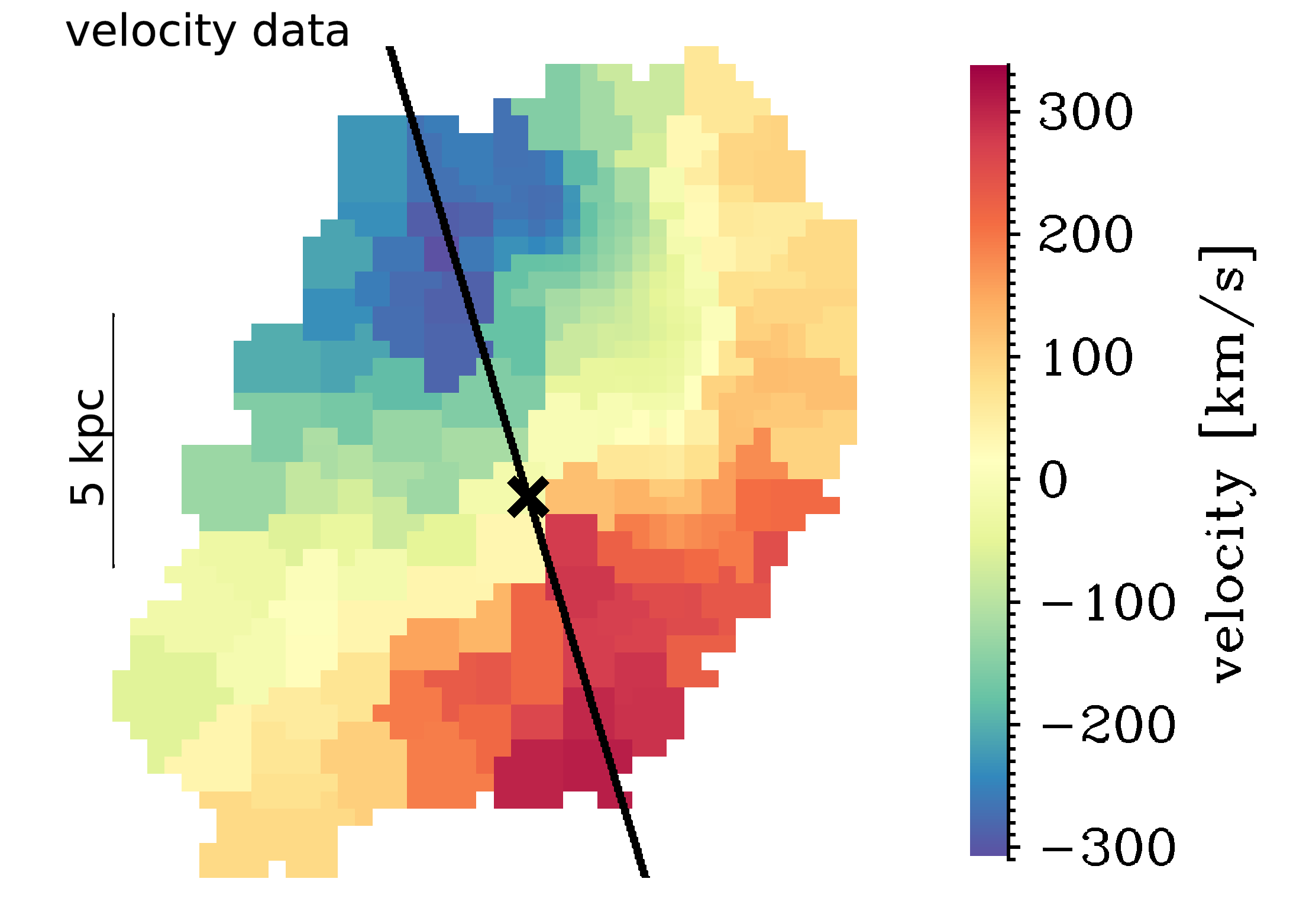}
    \includegraphics[width=0.24\textwidth]{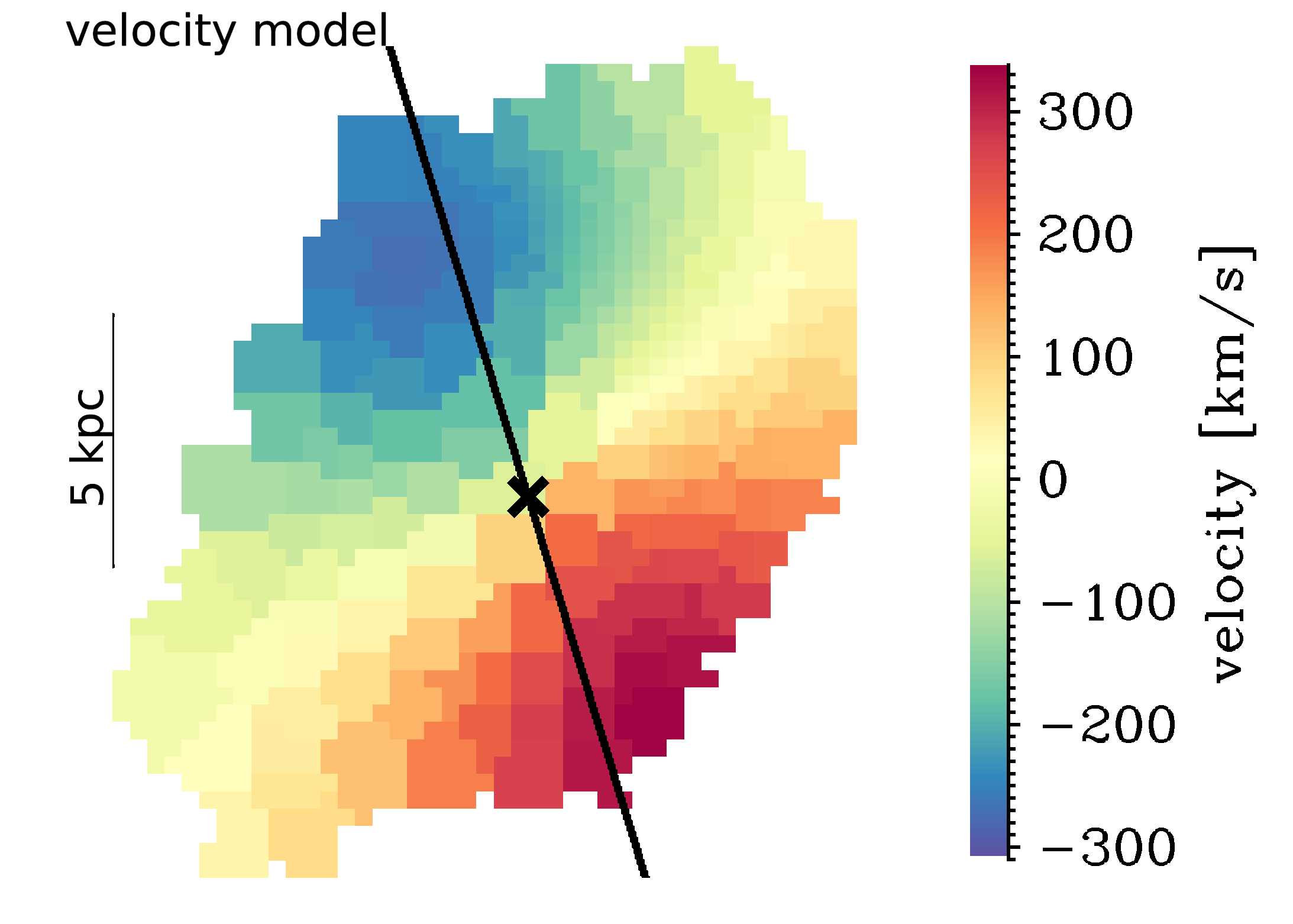}
    \includegraphics[width=0.24\textwidth]{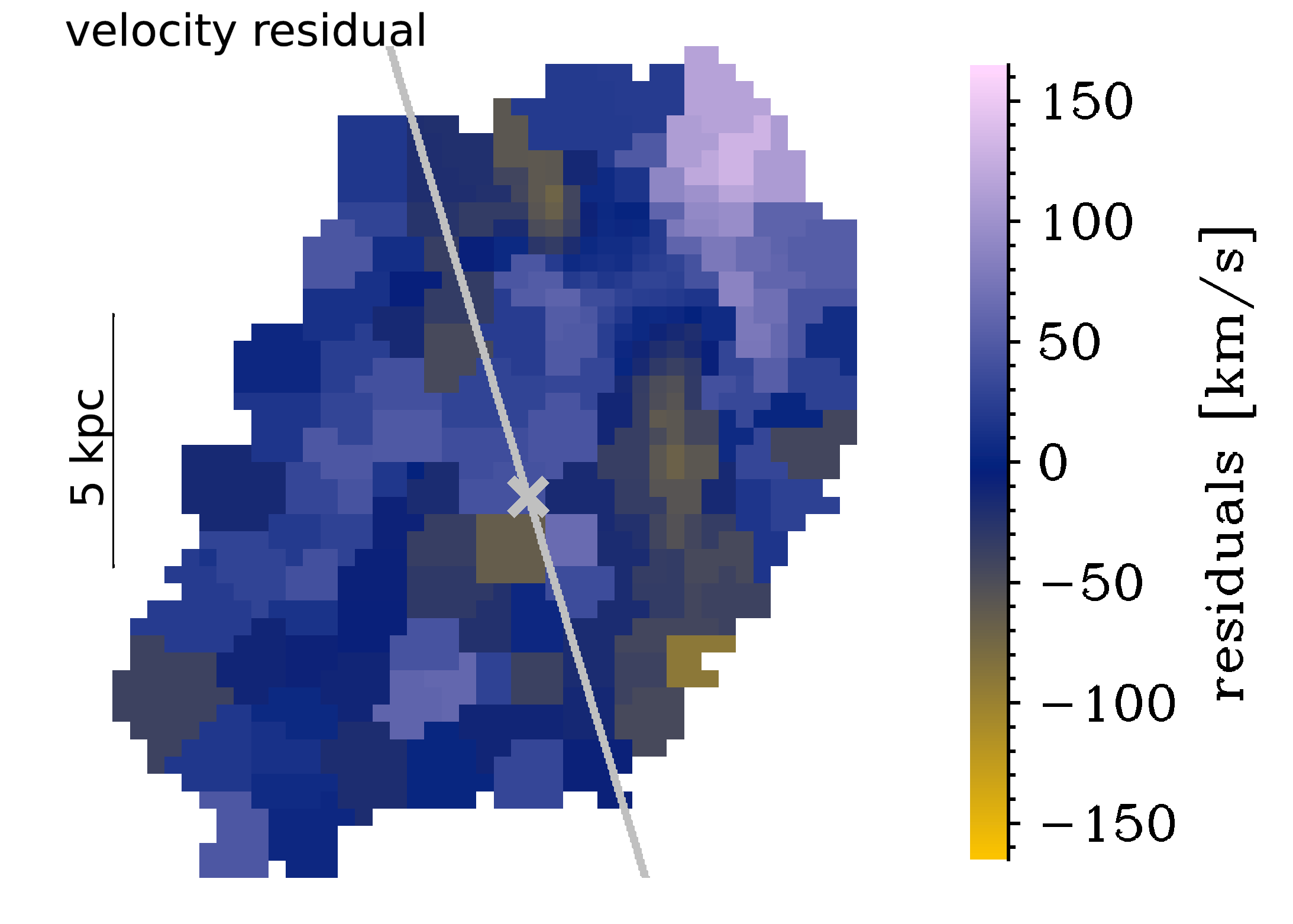}
    \includegraphics[width=0.24\textwidth]{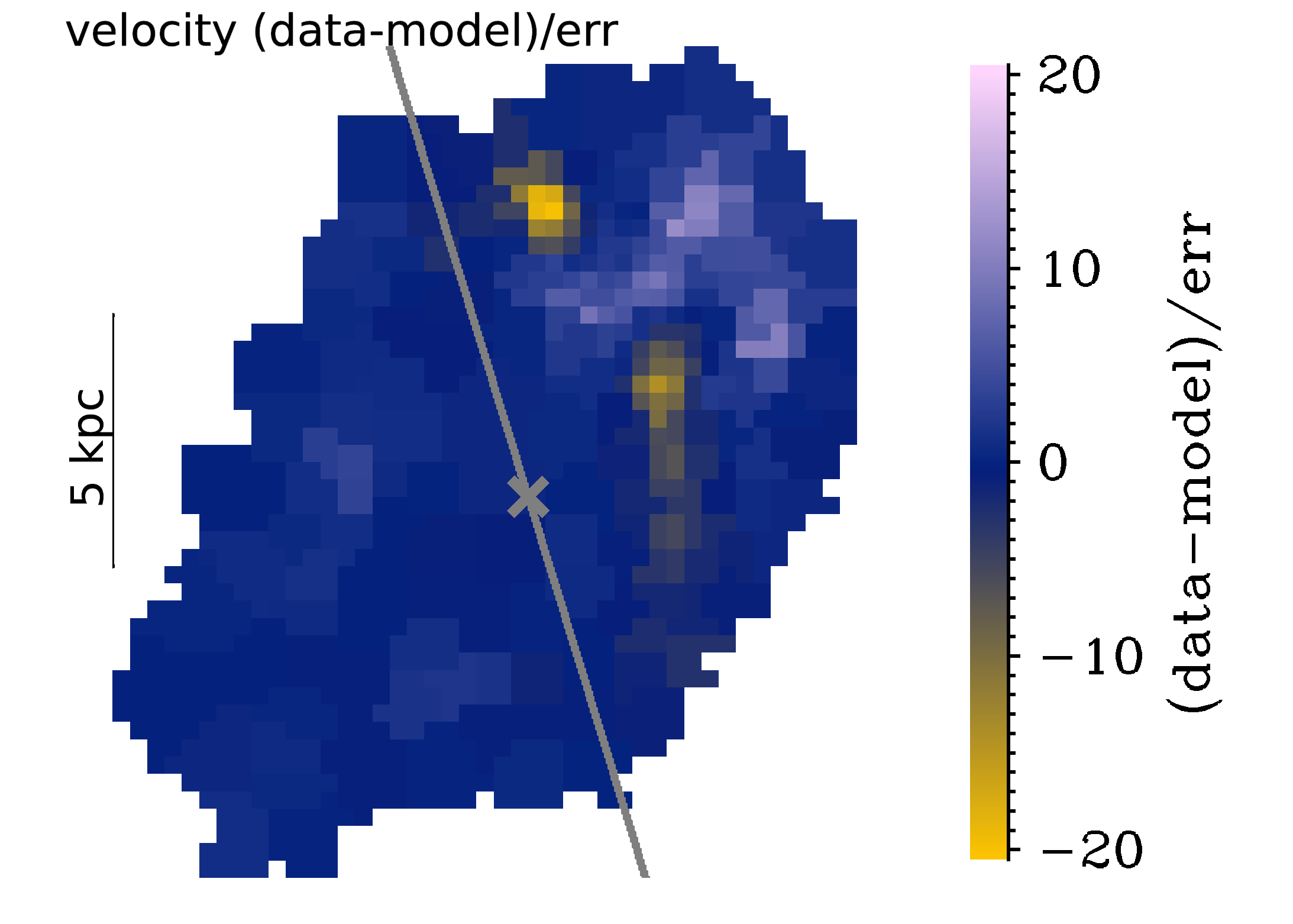}
    \includegraphics[width=0.24\textwidth]{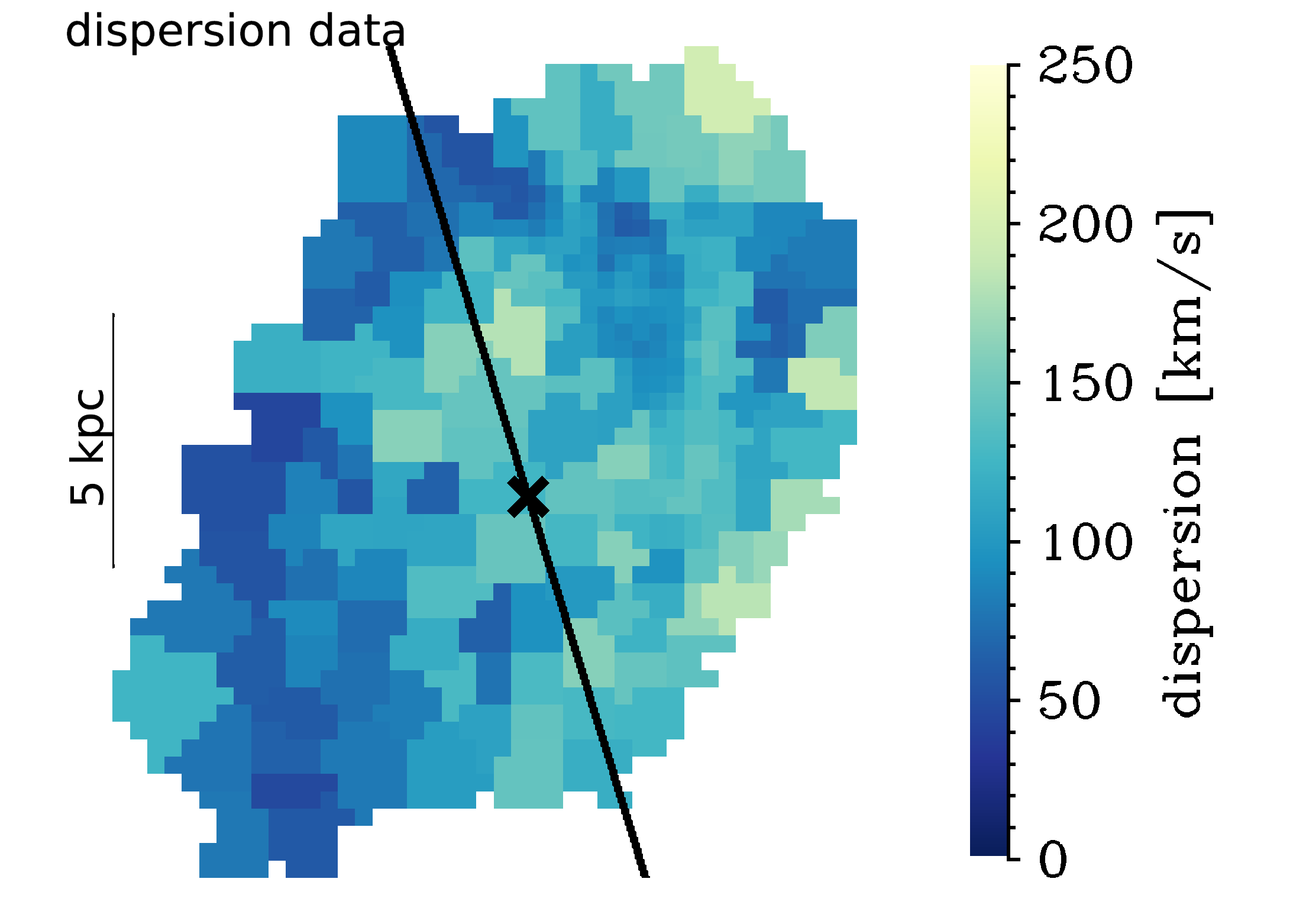}
    \includegraphics[width=0.24\textwidth]{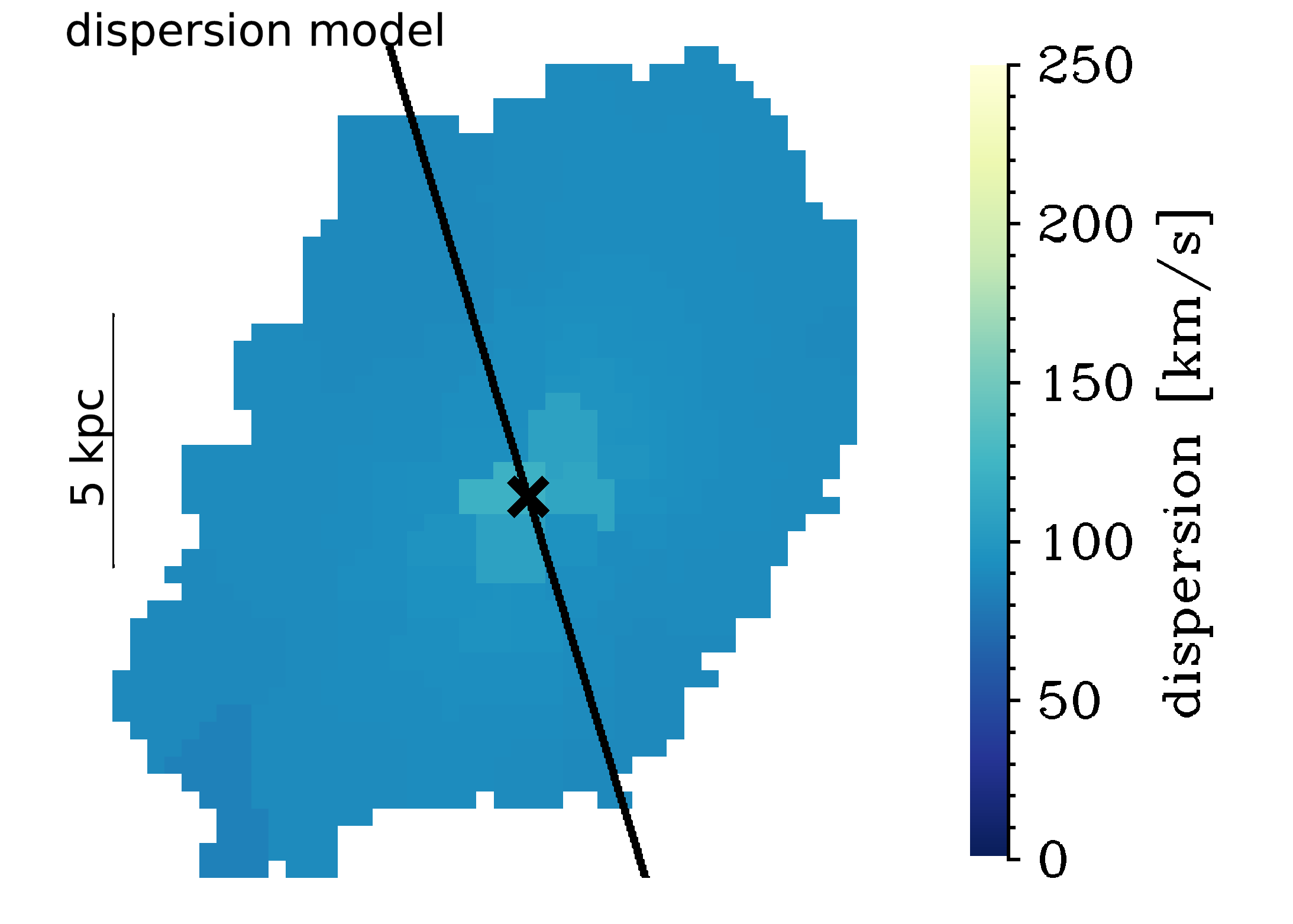}
    \includegraphics[width=0.24\textwidth]{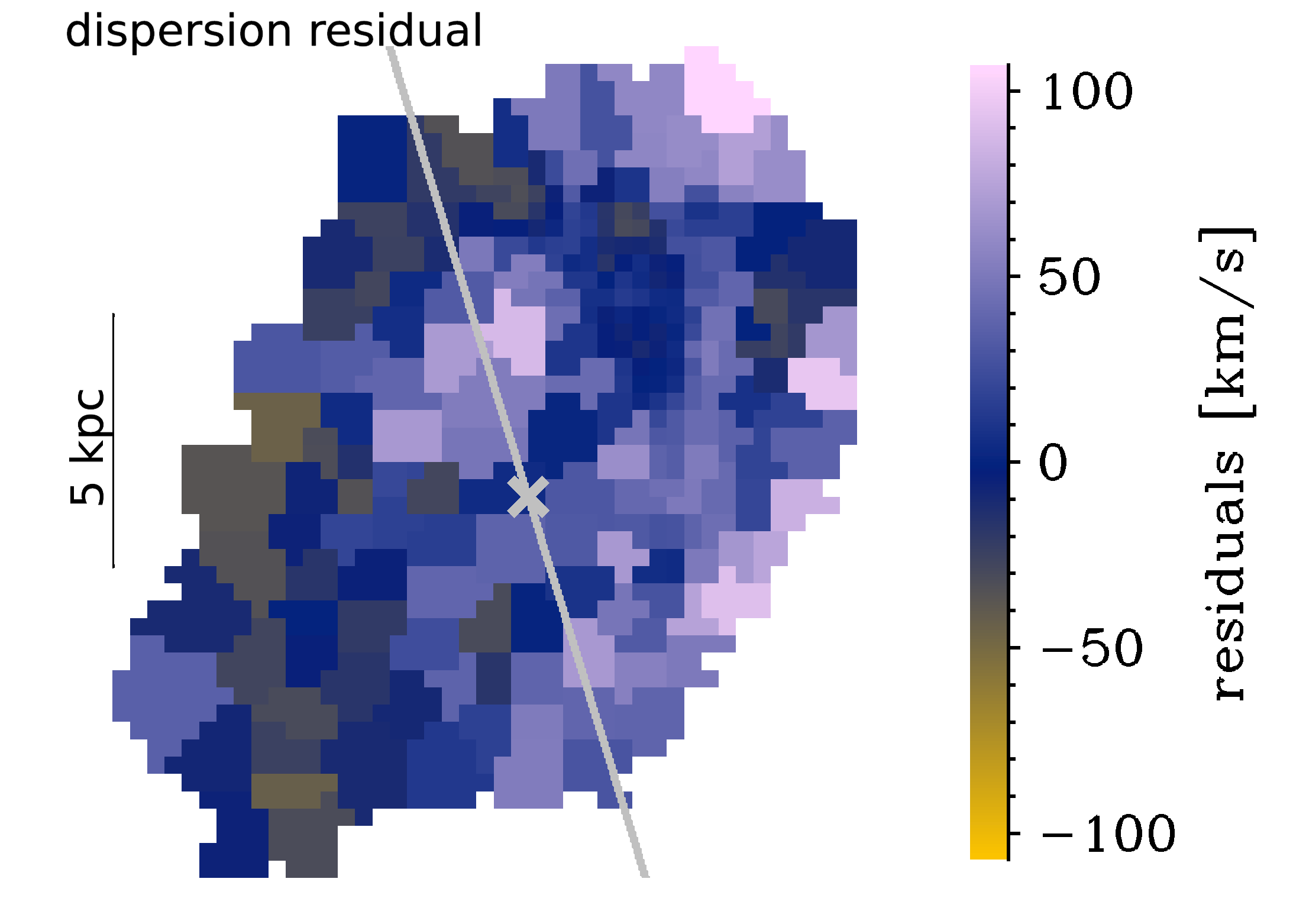}
    \includegraphics[width=0.24\textwidth]{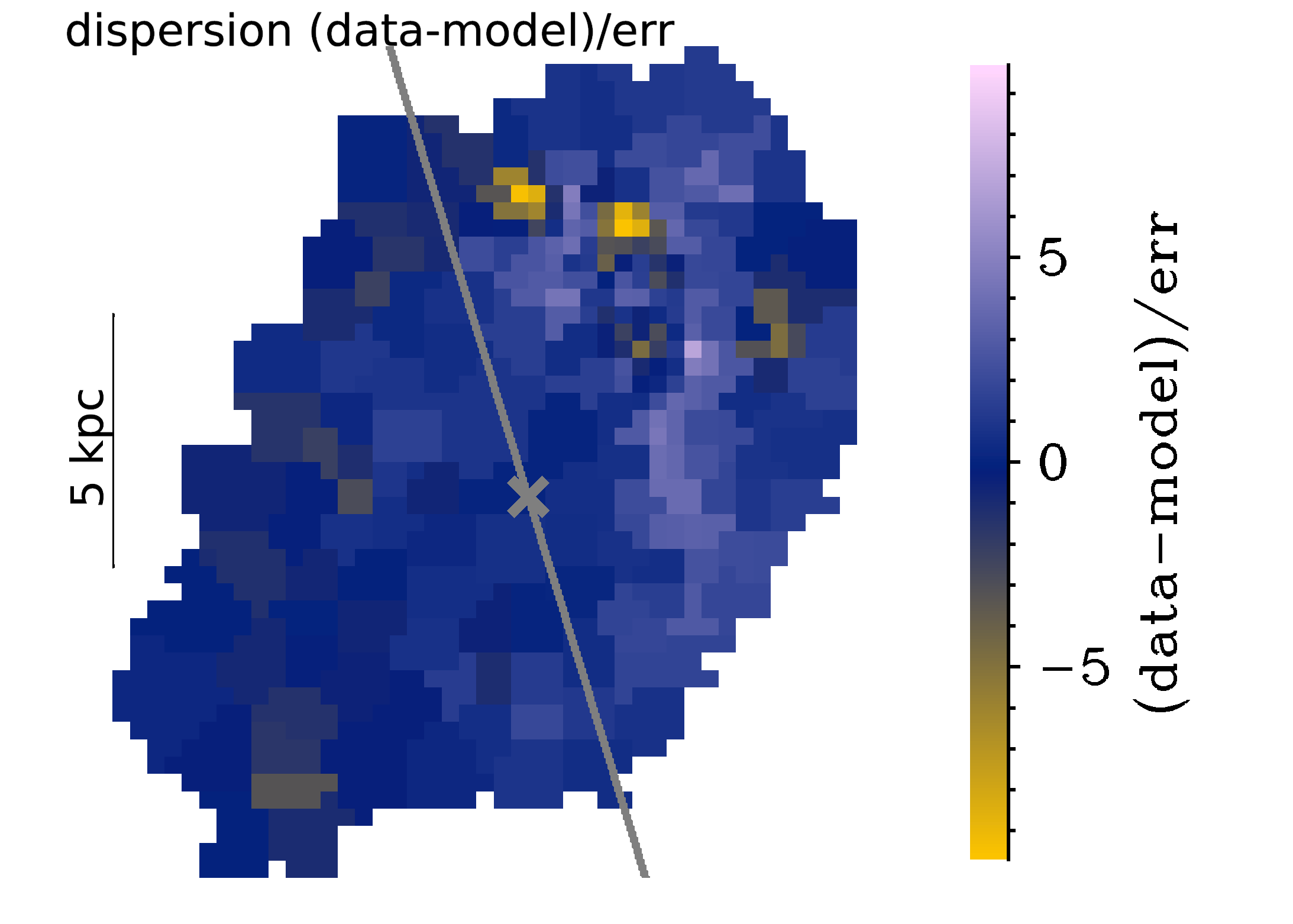}
    \caption{Same as Figure~\ref{f:model}, but including a planar radial inflow of $v_r=130$~km/s in addition to our fiducial disc, bulge and halo model. Top (bottom) panels show the observed velocity (velocity dispersion) field (left), the best-fit model (middle left), residuals (data-model; middle right), and goodness-of-fit (data-model)/uncertainties (right). The cross and line indicate the centre and best-fit PA. The median velocity offset is $\Delta v_{\rm med}=8.8$~km/s with a rms velocity difference of $\Delta v_{\rm rms}=40.0$~km/s. Corresponding values for the velocity dispersion are $\Delta\sigma_{\rm med}=15.4$~km/s and $\Delta\sigma_{\rm rms}=34.6$~km/s. While we observe strong residuals particularly in the North-Western region, as in our fiducial model, the magnitude of the residuals in the velocity field has decreased in this model, indicating that part of the deviations from circular motions could be explained by an inflow component.}
    \label{f:inflow}
\end{figure*}

In Figure~\ref{f:inflow} we show the results of our dynamical modelling including a uniform, planar radial inflow of fixed velocity $v_r=130$~km/s. I.e., we assume that gas is inflowing through the disc in an axisymmetric fashion and with radially constant velocity. The inflow velocity is not fitted for, but added as a fixed parameter \citep[see][]{Price21}. The preferred inflow velocity is initially identified through a grid search by injecting different values of $v_r$ in steps of 10~km/s from zero to 200~km/s, using least-squares minimization. We then obtain the best-fit through MCMC techniques as described in Section~\ref{s:fitting}, with the same free parameters as our fiducial model plus the radial inflow fixed at $v_r=130$~km/s.
Our modelling results are listed in Table~\ref{t:inflow}.

While we still find strong residuals with this second model specifically in the North-Western region of GN20, the overall magnitude of the velocity residuals decreases. The reduced chi-squared statistics are improved compared to our fiducial model presented in Section~\ref{s:kinematics} by $\Delta\chi^2_{\rm red}=0.33$. This shows that the inclusion of a radial inflow can help to explain some of the non-circular motions observed in GN20. 

We point out the change in the kinematic position angle of the best-fit model including inflow with respect to our fiducial model. The best-fit rotation velocities are reduced. In this model, we find little contribution of dark matter to the dynamics on galactic scales.

We compare the inflow velocity identified through our grid search to analytical estimates, as recently presented by \cite{Genzel23}. We follow their Equation~11 and parameter choices to obtain an analytical expectation for the inflow velocity. For this, we evaluate the Toomre$-Q$ parameter \citep{Toomre64} following \cite{BT08, Escala08, Dekel09b} at the radius where our best-fit circular velocity curve reaches its maximum \citep[see][]{Uebler19}, and find $Q_{\rm gas}\sim0.49$. Following \cite{Genzel23}, we then obtain $v_r=102$~km/s, which is somewhat lower than our model-derived value. 
If we approximate the inflow velocity through $v_r\sim f_{\rm gas}^2 v_c$ \citep{Genzel23}, with $v_c$ measured at the effective radius, we obtain $v_r=138$~km/s, comparable to our model-derived value.
In general, as pointed out by \cite{Genzel23}, analytical estimates of inflow velocities can vary by factors of a few \citep[e.g.][]{Dekel09b, Dekel13, Krumholz10, Krumholz18}. 
Our comparison to analytical estimates shows that the model-derived inflow velocity is not unexpected for a galaxy with the mass and kinematics of GN20.

Evidence for inflows in massive, high-redshift ($z\sim1-2.5$) star-forming disc galaxies has been found in deep ground-based observations \citep{Price21, Genzel23}. The radial velocities in those cases are measured to be $v_r\sim30-120$~km/s, comparable to the putative inflow velocity of $v_r=130$~km/s in GN20.

\begin{table}
\caption{Results from our best-fit dynamical model including a thick disc, bulge and dark matter halo, and a planar radial inflow of velocity $v_r=130$~km/s. For $\sigma_0$, we adopt minimum uncertainties of 10~km/s, and for the total baryonic mass we adopt minimum uncertainties of 0.1~dex. For $f_{\rm DM}$ we give the $3\sigma$ upper limit.}
\begin{tabular}{lc}
\toprule
    \multicolumn{2}{c}{Fitted parameters}\\
    $\log(M_{\rm bar}/M_\odot)$ & $11.5\pm0.1$ \\
    $\sigma_0$ [km/s] & $89\pm10$ \\
    $f_{\rm DM}(<R_e)$ & $<0.07$ \\
    PA$_{\rm kin}$ & $17.1^{+0.4}_{-0.5}$ \\
    velocity shift [km/s] & $31\pm2$ \\
    \hline
    $\chi^2_{\rm red}$ & 4.6414 \\  
    \hline
    \multicolumn{2}{c}{Derived values}\\
    $v_{\rm rot}(R_e)$ [km/s] & 443 \\
    $v_{\rm circ}(R_e)$ [km/s] & 472 \\
    $v_{\rm rot,max}$ [km/s] & 483 \\
    $v_{\rm rot(R_e)}/\sigma_0$ & $5.0\pm0.6$ \\
    $\log(M_{\rm dyn}(<2R_e)/M_\odot)$ & 11.51 \\
      
\end{tabular} 
\footnotesize{}
\label{t:inflow}
\end{table}

\begin{figure}
    \centering
    \includegraphics[width=0.7\columnwidth]{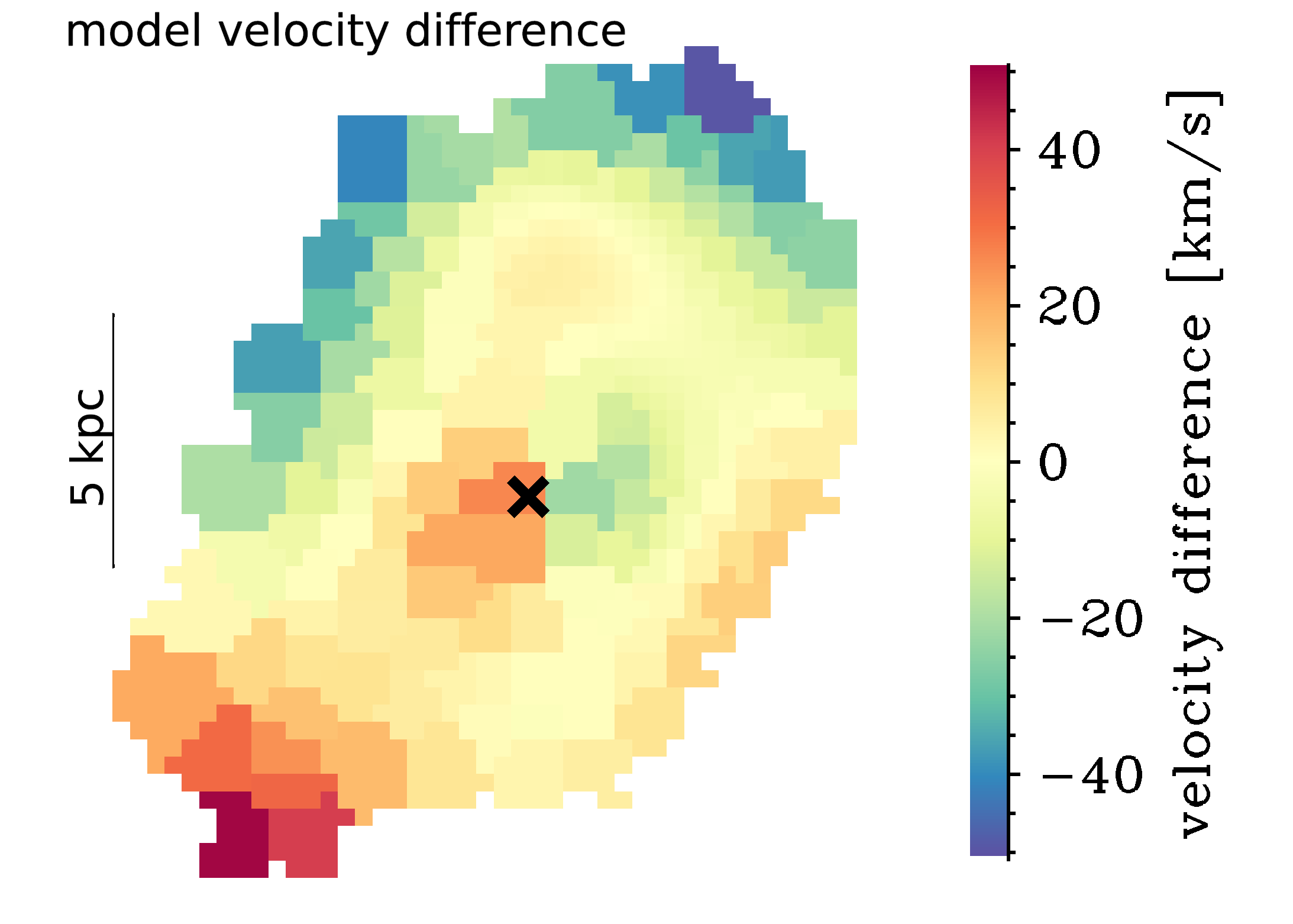}
    \caption{Difference in the model velocities of our fiducial model compared to the model including the radial inflow of $v_r=130$~km/s. The differences are most pronounced in the centre and in the outer regions.}
    \label{f:modelcomp}
\end{figure}

In Figure~\ref{f:modelcomp}, we show the difference between our fiducial velocity model as presented in the top middle-left panel of Figure~\ref{f:kinmaps}, and the velocity model including the radial inflow (top middle-left panel of Figure~\ref{f:inflow}). The differences between the two models are most pronounced in the centre and in the outer regions.

%%%%%%%%%%%%%%%%%%%%%%%%%%%%%%%%%%%%%%%%%%%%%%%%%%

% Don't change these lines
\bsp	% typesetting comment
\label{lastpage}
\end{document}